\pdfoutput=1
\documentclass{ar-1col}

\usepackage[T1]{fontenc}

\usepackage{silence}
\usepackage{ifthen} 

\newboolean{articletitles}
\setboolean{articletitles}{true} 
\newboolean{uprightparticles}
\setboolean{uprightparticles}{false} 

\def\paperauthors{Egede, Nishida, Patel, Schune}
\def\paperasciititle{Electroweak Penguin Decays of b-Flavoured Hadrons} 
\def\papertitle{Electroweak Penguin Decays of \bquark-Flavoured Hadrons} 
\def\paperkeywords{{BELLE-II}, {LHCb}, {Experimental}, {FCNC}} 
\def\papercopyright{\the\year\ The authors}

\def\paperlicenceurl{https://creativecommons.org/licenses/by/4.0/}

\usepackage{microtype}
\usepackage{lineno}  
\usepackage{xspace} 

\usepackage{graphicx}  
\graphicspath{{./figs/}} 

\usepackage{amsmath} 
\usepackage{amssymb}
\usepackage{amsfonts}
\usepackage{upgreek} 

\usepackage{bigints}

\newcommand*\patchAmsMathEnvironmentForLineno[1]{%
\expandafter\let\csname old#1\expandafter\endcsname\csname #1\endcsname
\expandafter\let\csname oldend#1\expandafter\endcsname\csname
end#1\endcsname
 \renewenvironment{#1}%
   {\linenomath\csname old#1\endcsname}%
   {\csname oldend#1\endcsname\endlinenomath}%
}
\newcommand*\patchBothAmsMathEnvironmentsForLineno[1]{%
  \patchAmsMathEnvironmentForLineno{#1}%
  \patchAmsMathEnvironmentForLineno{#1*}%
}
\AtBeginDocument{%
\patchBothAmsMathEnvironmentsForLineno{equation}%
\patchBothAmsMathEnvironmentsForLineno{align}%
\patchBothAmsMathEnvironmentsForLineno{flalign}%
\patchBothAmsMathEnvironmentsForLineno{alignat}%
\patchBothAmsMathEnvironmentsForLineno{gather}%
\patchBothAmsMathEnvironmentsForLineno{multline}%
\patchBothAmsMathEnvironmentsForLineno{eqnarray}%
}

\usepackage{hyperxmp}

\usepackage[pdftex,
            pdfauthor={\paperauthors},
            pdftitle={\paperasciititle},
            pdfkeywords={\paperkeywords},
            pdfcopyright={Copyright (C) \papercopyright},
            pdflicenseurl={\paperlicenceurl}]{hyperref}

\usepackage[all]{hypcap} 

\usepackage{onimage}
\def\linkRect(#1)(#2){
  \coordinate (ll) at (#1); \coordinate (ur) at (#2);\def~{\tikz \useasboundingbox (ll) rectangle (ur);}
  \node [anchor=south west, inner sep=0] at (ll)
}

\usepackage[flushleft]{threeparttable}

\usepackage{xspace} 
\usepackage{upgreek}

\def\lhcb   {\mbox{LHCb}\xspace}
\def\atlas  {\mbox{ATLAS}\xspace}
\def\cms    {\mbox{CMS}\xspace}

\def\babar  {\mbox{BaBar}\xspace}
\def\belle  {\mbox{Belle}\xspace}
\def\belletwo {\mbox{Belle~II}\xspace}

\def\cleo   {\mbox{CLEO}\xspace}

\def\lhc    {\mbox{LHC}\xspace}

\def\MagUp {\mbox{\em Mag\kern -0.05em Up}\xspace}

\ifthenelse{\boolean{uprightparticles}}%
{
 
 \def\Pgamma      {\ensuremath{\upgamma}\xspace}

 \def\Pmu         {\ensuremath{\upmu}\xspace}                 
 \def\Pnu         {\ensuremath{\upnu}\xspace}                 
                  
 \def\Ppi         {\ensuremath{\uppi}\xspace}

 \def\Ptau        {\ensuremath{\uptau}\xspace}

 \def\Ppsi        {\ensuremath{\uppsi}\xspace}

 \def\PDelta      {\ensuremath{\Delta}\xspace}                 
 \def\PXi         {\ensuremath{\Xi}\xspace}                 
 \def\PLambda     {\ensuremath{\Lambda}\xspace}                 
 \def\PSigma      {\ensuremath{\Sigma}\xspace}                 
 \def\POmega      {\ensuremath{\Omega}\xspace}                 
 \def\PUpsilon    {\ensuremath{\Upsilon}\xspace}

 \def\PB      {\ensuremath{\mathrm{B}}\xspace}                 
                  
 \def\PD      {\ensuremath{\mathrm{D}}\xspace}

 \def\PJ      {\ensuremath{\mathrm{J}}\xspace}                 
 \def\PK      {\ensuremath{\mathrm{K}}\xspace}

 \def\PW      {\ensuremath{\mathrm{W}}\xspace}

 \def\Pb      {\ensuremath{\mathrm{b}}\xspace}                 
 \def\Pc      {\ensuremath{\mathrm{c}}\xspace}                 
 \def\Pd      {\ensuremath{\mathrm{d}}\xspace}                 
 \def\Pe      {\ensuremath{\mathrm{e}}\xspace}

 \def\Pi      {\ensuremath{\mathrm{i}}\xspace}

 \def\Pq      {\ensuremath{\mathrm{q}}\xspace}                 
                  
 \def\Ps      {\ensuremath{\mathrm{s}}\xspace}                 
 \def\Pt      {\ensuremath{\mathrm{t}}\xspace}                 
 \def\Pu      {\ensuremath{\mathrm{u}}\xspace}

 \def\thebaroffset{0.0em}
}
{
 
 \def\Pgamma      {\ensuremath{\gamma}\xspace}

 \def\Pmu         {\ensuremath{\mu}\xspace}                 
 \def\Pnu         {\ensuremath{\nu}\xspace}                 
                  
 \def\Ppi         {\ensuremath{\pi}\xspace}

 \def\Ptau        {\ensuremath{\tau}\xspace}

 \def\Ppsi        {\ensuremath{\psi}\xspace}                 
                  
 \mathchardef\PDelta="7101
 \mathchardef\PXi="7104
 \mathchardef\PLambda="7103
 \mathchardef\PSigma="7106
 \mathchardef\POmega="710A
 \mathchardef\PUpsilon="7107
                  
 \def\PB      {\ensuremath{B}\xspace}                 
                  
 \def\PD      {\ensuremath{D}\xspace}

 \def\PJ      {\ensuremath{J}\xspace}                 
 \def\PK      {\ensuremath{K}\xspace}

 \def\PW      {\ensuremath{W}\xspace}

 \def\Pb      {\ensuremath{b}\xspace}                 
 \def\Pc      {\ensuremath{c}\xspace}                 
 \def\Pd      {\ensuremath{d}\xspace}                 
 \def\Pe      {\ensuremath{e}\xspace}

 \def\Pi      {\ensuremath{i}\xspace}

 \def\Pq      {\ensuremath{q}\xspace}                 
                  
 \def\Ps      {\ensuremath{s}\xspace}                 
 \def\Pt      {\ensuremath{t}\xspace}                 
 \def\Pu      {\ensuremath{u}\xspace}

 \def\thebaroffset{0.18em}
}
\newcommand{\offsetoverline}[2][\thebaroffset]{\kern #1\overline{\kern -#1 #2}}%

\makeatletter
\newcommand{\miniscule}{\@setfontsize\miniscule{5}{6}}
\makeatother

\DeclareRobustCommand{\optbar}[1]{\shortstack{{\miniscule (\rule[.5ex]{1.25em}{.18mm})}
  \\ [-.7ex] $#1$}}

\def\en         {{\ensuremath{\Pe^-}}\xspace}   
\def\ep         {{\ensuremath{\Pe^+}}\xspace}

\def\epem       {{\ensuremath{\Pe^+\Pe^-}}\xspace}

\def\mup        {{\ensuremath{\Pmu^+}}\xspace}
\def\mun        {{\ensuremath{\Pmu^-}}\xspace} 

\def\mumu       {{\ensuremath{\Pmu^+\Pmu^-}}\xspace}

\def\taup       {{\ensuremath{\Ptau^+}}\xspace}
\def\taum       {{\ensuremath{\Ptau^-}}\xspace}

\def\ellell     {\ensuremath{\ell^+ \ell^-}\xspace}

\def\neu        {{\ensuremath{\Pnu}}\xspace}
\def\neub       {{\ensuremath{\overline{\Pnu}}}\xspace}

\def\neumb      {{\ensuremath{\neub_\mu}}\xspace}

\def\g      {{\ensuremath{\Pgamma}}\xspace}

\def\W      {{\ensuremath{\PW}}\xspace}

\def\quark     {{\ensuremath{\Pq}}\xspace}
\def\quarkbar  {{\ensuremath{\overline \quark}}\xspace}
\def\qqbar     {{\ensuremath{\quark\quarkbar}}\xspace}
\def\uquark    {{\ensuremath{\Pu}}\xspace}

\def\dquark    {{\ensuremath{\Pd}}\xspace}

\def\squark    {{\ensuremath{\Ps}}\xspace}
\def\squarkbar {{\ensuremath{\overline \squark}}\xspace}

\def\cquark    {{\ensuremath{\Pc}}\xspace}
\def\cquarkbar {{\ensuremath{\overline \cquark}}\xspace}

\def\bquark    {{\ensuremath{\Pb}}\xspace}
\def\bquarkbar {{\ensuremath{\overline \bquark}}\xspace}
\def\bbbar     {{\ensuremath{\bquark\bquarkbar}}\xspace}
\def\tquark    {{\ensuremath{\Pt}}\xspace}

\def\pion   {{\ensuremath{\Ppi}}\xspace}
\def\piz    {{\ensuremath{\pion^0}}\xspace}
\def\pip    {{\ensuremath{\pion^+}}\xspace}
\def\pim    {{\ensuremath{\pion^-}}\xspace}

\def\kaon    {{\ensuremath{\PK}}\xspace}
\def\Kbar    {{\ensuremath{\offsetoverline{\PK}}}\xspace}

\def\KorKbar {\kern \thebaroffset\optbar{\kern -\thebaroffset \PK}{}\xspace}

\def\Kp      {{\ensuremath{\kaon^+}}\xspace}
\def\Km      {{\ensuremath{\kaon^-}}\xspace}

\def\KS      {{\ensuremath{\kaon^0_{\mathrm{S}}}}\xspace}

\def\Kstarz  {{\ensuremath{\kaon^{*0}}}\xspace}
\def\Kstarzb {{\ensuremath{\Kbar{}^{*0}}}\xspace}

\def\Kstarb  {{\ensuremath{\Kbar{}^*}}\xspace}

\def\Kstarm  {{\ensuremath{\kaon^{*-}}}\xspace}

\def\D       {{\ensuremath{\PD}}\xspace}

\def\DorDbar {\kern \thebaroffset\optbar{\kern -\thebaroffset \PD}\xspace}
\def\Dz      {{\ensuremath{\D^0}}\xspace}

\def\Dp      {{\ensuremath{\D^+}}\xspace}
\def\Dm      {{\ensuremath{\D^-}}\xspace}

\def\DpDm    {\ensuremath{\Dp {\kern -0.16em \Dm}}\xspace}

\def\Dstarp  {{\ensuremath{\D^{*+}}}\xspace}

\def\B       {{\ensuremath{\PB}}\xspace}
\def\Bbar    {{\ensuremath{\offsetoverline{\PB}}}\xspace}
\def\Bb      {{\ensuremath{\Bbar}}\xspace}
\def\BorBbar {\kern \thebaroffset\optbar{\kern -\thebaroffset \PB}\xspace}
\def\Bz      {{\ensuremath{\B^0}}\xspace}
\def\Bzb     {{\ensuremath{\Bbar{}^0}}\xspace}
\def\Bd      {{\ensuremath{\B^0}}\xspace}
\def\Bdb     {{\ensuremath{\Bbar{}^0}}\xspace}
\def\BdorBdbar {\kern \thebaroffset\optbar{\kern -\thebaroffset \Bd}\xspace}
\def\Bu      {{\ensuremath{\B^+}}\xspace}
\def\Bub     {{\ensuremath{\B^-}}\xspace}
\def\Bp      {{\ensuremath{\Bu}}\xspace}
\def\Bm      {{\ensuremath{\Bub}}\xspace}

\def\Bs      {{\ensuremath{\B^0_\squark}}\xspace}
\def\Bsb     {{\ensuremath{\Bbar{}^0_\squark}}\xspace}
\def\BsorBsbar {\kern \thebaroffset\optbar{\kern -\thebaroffset \Bs}\xspace}
\def\Bc      {{\ensuremath{\B_\cquark^+}}\xspace}

\def\Bcm     {{\ensuremath{\B_\cquark^-}}\xspace}

\def\jpsi     {{\ensuremath{{\PJ\mskip -3mu/\mskip -2mu\Ppsi}}}\xspace}
\def\psitwos  {{\ensuremath{\Ppsi{(2S)}}}\xspace}

\def\Y#1S{\ensuremath{\PUpsilon{(#1S)}}\xspace}

\def\FourS {{\Y4S}}
\def\FiveS {{\Y5S}}

\def\Lz          {{\ensuremath{\PLambda}}\xspace}

\def\LorLbar     {\kern \thebaroffset\optbar{\kern -\thebaroffset \PLambda}\xspace}

\def\Lb           {{\ensuremath{\Lz^0_\bquark}}\xspace}

\newcommand{\decay}[2]{\ensuremath{#1\!\to #2}\xspace} 

\def\to                 {\ensuremath{\rightarrow}\xspace}

\def\qsq       {{\ensuremath{q^2}}\xspace}

\def\CP                {{\ensuremath{C\!P}}\xspace}

\def\Vud  {{\ensuremath{V_{\uquark\dquark}^{\phantom{\ast}}}}\xspace}
\def\Vcd  {{\ensuremath{V_{\cquark\dquark}^{\phantom{\ast}}}}\xspace}
\def\Vtd  {{\ensuremath{V_{\tquark\dquark}^{\phantom{\ast}}}}\xspace}

\def\Vts  {{\ensuremath{V_{\tquark\squark}^{\phantom{\ast}}}}\xspace}
\def\Vub  {{\ensuremath{V_{\uquark\bquark}^{\phantom{\ast}}}}\xspace}
\def\Vcb  {{\ensuremath{V_{\cquark\bquark}^{\phantom{\ast}}}}\xspace}
\def\Vtb  {{\ensuremath{V_{\tquark\bquark}^{\phantom{\ast}}}}\xspace}

\def\Vtss  {{\ensuremath{V_{\tquark\squark}^\ast}}\xspace}

\newcommand{\ACP}{{\ensuremath{{\mathcal{A}}^{\CP}}}\xspace}

\def\bsll     {\decay{\bquark}{\squark \ell^+ \ell^-}}

\def\AT#1     {\ensuremath{A_{\mathrm{T}}^{#1}}\xspace}           
\def\btosgam  {\decay{\bquark}{\squark \g}}
\def\btodgam  {\decay{\bquark}{\dquark \g}}

\def\C#1      {\ensuremath{\mathcal{C}_{#1}}\xspace}                       
\def\Cp#1     {\ensuremath{\mathcal{C}_{#1}^{'}}\xspace}                    
\def\Ceff#1   {\ensuremath{\mathcal{C}_{#1}^{\mathrm{(eff)}}}\xspace}        
\def\Cpeff#1  {\ensuremath{\mathcal{C}_{#1}^{'\mathrm{(eff)}}}\xspace}       
\def\Ope#1    {\ensuremath{\mathcal{O}_{#1}}\xspace}                       
\def\Opep#1   {\ensuremath{\mathcal{O}_{#1}^{'}}\xspace}                    

\newcommand{\aunit}[1]{\ensuremath{\text{\,#1}}}       

\newcommand{\tev}{\aunit{Te\kern -0.1em V}\xspace}
\newcommand{\gev}{\aunit{Ge\kern -0.1em V}\xspace}
\newcommand{\mev}{\aunit{Me\kern -0.1em V}\xspace}
\newcommand{\kev}{\aunit{ke\kern -0.1em V}\xspace}
\newcommand{\ev}{\aunit{e\kern -0.1em V}\xspace}
 
\newcommand{\mevc}{\ensuremath{\aunit{Me\kern -0.1em V\!/}c}\xspace}
\newcommand{\gevc}{\ensuremath{\aunit{Ge\kern -0.1em V\!/}c}\xspace}
\newcommand{\mevcc}{\ensuremath{\aunit{Me\kern -0.1em V\!/}c^2}\xspace}
\newcommand{\gevcc}{\ensuremath{\aunit{Ge\kern -0.1em V\!/}c^2}\xspace}
\newcommand{\gevgevcccc}{\ensuremath{\gev^2\!/c^4}\xspace} 

\def\cm   {\aunit{cm}\xspace}

\def\nb {\aunit{nb}\xspace}

\def\fb   {\ensuremath{\aunit{fb}}\xspace}
\def\invfb   {\ensuremath{\fb^{-1}}\xspace}
\def\ab   {\ensuremath{\aunit{ab}}\xspace}
\def\invab   {\ensuremath{\ab^{-1}}\xspace}

\def\gsim{{~\raise.15em\hbox{$>$}\kern-.85em
          \lower.35em\hbox{$\sim$}~}\xspace}
\def\lsim{{~\raise.15em\hbox{$<$}\kern-.85em
          \lower.35em\hbox{$\sim$}~}\xspace}

\def\tell1  {TELL1\xspace}
\def\ukl1   {UKL1\xspace}

\newcommand{\eg}{\mbox{\itshape e.g.}\xspace}
\newcommand{\ie}{\mbox{\itshape i.e.}\xspace}

\def\btosll {\decay{\bquark}{\squark\ellell}}

\def\BdToKstKpiJPsll  {\decay{\Bd}{\jpsi (\rightarrow \ellell )\Kstarz (\rightarrow \Kp \pim)}}

\def\bTosmm{\decay{\bquark}{\squark \, \mu^+ \mu^-}}
\def\bTosee{\decay{\bquark}{\squark \, e^+ e^-}}
\def\bTosll{\decay{\bquark}{\squark \, \ell^+ \ell^-}}
\def\qsqmin{\ensuremath{q^2_{\mathrm{min}}}\xspace}
\def\qsqmax{\ensuremath{q^2_{\mathrm{max}}}\xspace}
\def\lplm{\ell^+ \ell^-}

\def\RK{\ensuremath{R_{\kaon}}\xspace}

\usepackage{cite} 
\usepackage{mciteplus}

\newcommand{\PMerr}[2]{{\,}^{+\,#1}_{-\,#2}}

\begin{document}

\markboth{Egede et al.}{\papertitle}

\title{\papertitle}

\author{Ulrik Egede,$^1$ Shohei Nishida,$^{2,3}$ Mitesh Patel,$^4$ Marie-H\'{e}l\`{e}ne Schune$^5$ 
\affil{$^1$School of Physics \& Astronomy, Monash University, Melbourne, Australia; email: ulrik.egede@monash.edu}
\affil{$^2$ The Graduate University for Advanced Studies (SOKENDAI), Hayama, Japan}
\affil{$^3$ High Energy Accelerator Research Organization (KEK), Tsukuba, Japan; 
 email:shohei.nishida@kek.jp}
\affil{$^4$Department of Physics, Imperial College London, London, United Kingdom; email: mitesh.patel@imperial.ac.uk}
\affil{$^5$ Universit\'{e} Paris-Saclay, CNRS/IN2P3, IJCLab, Orsay, France; 
 email:marie-helene.schune@ijclab.in2p3.fr}}

\begin{abstract}
In the past decade, electroweak penguin decays have provided a number of precision measurements, turning into one of the most competitive ways to search for New Physics that describe beyond the Standard Model phenomena. An overview of the measurements made at the \B factories and hadron colliders are given and the experimental methods are presented. Experimental measurements required to provide further insight into present indications of New Physics are discussed.
\end{abstract}

\maketitle

\tableofcontents

\section{INTRODUCTION}
\label{sec:Introduction}
\begin{figure}[b]
\centering
    \includegraphics[width=1.0\columnwidth]{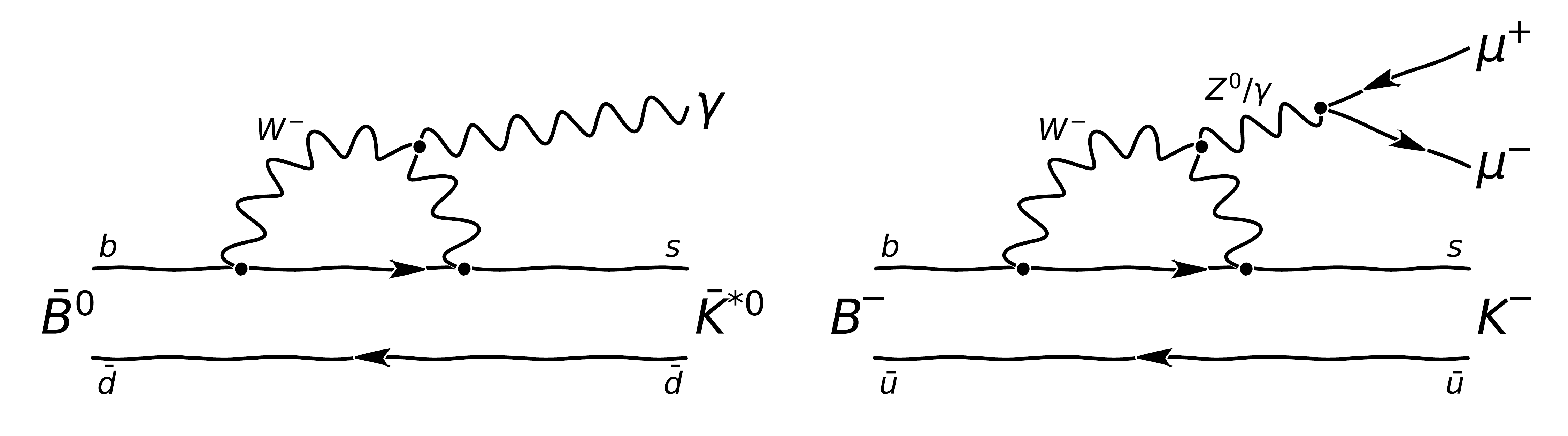}
    \caption{Example Feynman diagrams for the flavour changing neutral current~(FCNC) decays (left) \decay{\Bzb}{\Kstarzb\gamma} and (right) \decay{\Bm}{\Km\mup\mun} in the Standard Model.}
    \label{fig:FCNC}
\end{figure}

The most common decays of \bquark-hadrons take place at the quark level through the decay of the \bquark quark via the emission of a virtual \W boson. The \decay{\bquark}{\cquark\W} process mediates decays like \decay{\Bm}{\Dz\pim}, and the \decay{\bquark}{\uquark\W} process mediates decays like \decay{\Bzb}{\pip\ell^-\neub}. The \bquark quark is not able to decay directly to an \squark quark as it would require a vertex with a neutral vector boson and a change of flavour. Such a \emph{Flavour Changing Neutral Current}~(FCNC) process is forbidden in the Standard Model~(SM) at tree level. However, at one-loop level the FCNC quark-level process like \decay{\bquark}{\squark\gamma} is allowed in the SM as illustrated for the decays \decay{\Bzb}{\Kstarzb\gamma} and \decay{\Bm}{\Km\mup\mun} in Fig.~\ref{fig:FCNC}. Collectively, decays of this type with a hard photon or a lepton pair in the final state are known as \emph{electroweak penguin decays}. In the SM, the GIM mechanism~\cite{Glashow:1970gm} would, in the limit where the quarks inside the loop have zero mass compared to the \W boson mass, result in a cancellation of the decay amplitude to all orders. However, the large mass of the top quark means this is not the case for \bquark-hadron decays. On the other hand, electroweak penguin decays of \cquark-hadrons are very heavily suppressed in the SM. 
\begin{marginnote}[]
\entry{FCNC}{A Flavour Changing Neural Current~(FCNC) is a process where the quark involved changes flavour but keeps the same charge. An example is the \decay{\bquark}{\squark\gamma} process. At tree level the process is forbidden in the SM.} 
\end{marginnote}
\begin{marginnote}[]
\entry{Penguin decay}{The class of FCNC decays that are mediated by one loop Feynman diagrams. The name is related to that the diagrams can be drawn in the shape of penguins~\cite{Shifman:1995hc}.} 
\end{marginnote}

\subsection{Discovery modes for New Physics}
Electroweak penguin decays can act as a discovery mode for New Physics~(NP). At either tree level or at loop level, there can be particles such as new vector bosons or leptoquarks that mediate the decays. The influence of these particles can be observed as differences with respect to the SM predictions for these decays. While this statement is in principle true for the decay of any hadron, the study of electroweak penguin decays are very well suited as a discovery mode as: the SM amplitudes are suppressed due to the requirement of a loop-level process, thus making any NP effect more visible; the theoretical calculation of the final-state properties has lower uncertainties than in fully-hadronic decays; and final states with leptons or hard photons are relative easy to identify in particle physics detectors.

\subsection{Theoretical framework}

The theoretical calculation of the electroweak penguin decays uses an Operator Product Expansion~(OPE)~\cite{Mannel:2004ce}. The principle is the same as for the Fermi-theory of weak decays. It takes advantage of the fact that the decays are only sensitive to the spin, parity and \CP properties of the couplings involving particles at masses well above the \bquark-hadron masses. For processes like \decay{\bquark}{\squark\gamma}, \decay{\bquark}{\squark\ellell} and \decay{\bquark}{\squark\neu\neub}, it is possible within the OPE to write
\begin{equation}
    \label{eq:Heff}
    \mathcal{H}_{\rm{eff}} = \frac{4G_F}{\sqrt{2}} \Vtb\Vtss 
    \hspace{-1em} \sum_{i=7,9,10,\nu}\hspace{-1em}{\left[ C_i \mathcal{O}_i + C'_i \mathcal{O}'_i \right]} \; ,
\end{equation}
where the operators $\mathcal{O}_i$ encode the low energy behaviour and the complex valued Wilson coefficients $C_i$ characterise how these different operators contribute to the overall processes. The electromagnetic operator is $\mathcal{O}_7$, the semileptonic operators $\mathcal{O}_9$ and $\mathcal{O}_{10}$ correspond to vector and axial-vector currents, and $\mathcal{O}_\neu$ is the operator corresponding to \decay{\bquark}{\squark\neu\neub} processes. The QCD operators $\mathcal{O}_{\text{1--6,8}}$ mix through renormalisation with the operators $\mathcal{O}_{7,9}$, leading to measurements that are sensitive to the effective Wilson coefficients $C^{\rm{eff}}_{7,9}$. 
The primed operators correspond to right-handed currents and are suppressed by a factor $m_\squark/m_\bquark$ in the SM. Any NP will manifest itself through Wilson coefficients that have different values from those expected in the SM or through Wilson coefficients that correspond to completely new operators such as scalar, pseudoscalar or tensor currents. In the SM, the coupling to all leptons is the same and the validity of this lepton flavour universality can be probed by testing whether the Wilson coefficients have the same values for different flavours of lepton pairs in the final state. Contributions to the Wilson coefficients from NP depend on both the coupling constants between the NP particles and the SM particles and the masses of the NP particles. This means that the study of electroweak penguin decays cannot determine precisely the mass of any NP particles.

The branching fraction of a specific decay and the angular distribution of the decay products depends not only on the physics at the high energy scale described by the effective Hamiltonian. The dominant effect of the hadrons in the final state are described through \qsq--dependent form factors, where \qsq is the mass squared of the lepton pair. These describe low-energy QCD effects and as such cannot be calculated using perturbative methods. Rather, light cone sum rule calculations are used at low \qsq and lattice QCD at high \qsq. The uncertainties on the form factors are significant and measurements can be divided up into kinematic regions, that have different dependence on non-perturbative effects.
\begin{marginnote}[]
\entry{\qsq}{The \emph{q-squared} of an electroweak penguin decay is the mass squared of the lepton pair in the final state.} 
\end{marginnote}

\begin{figure}
    \centering
    \includegraphics[width=0.95\linewidth]{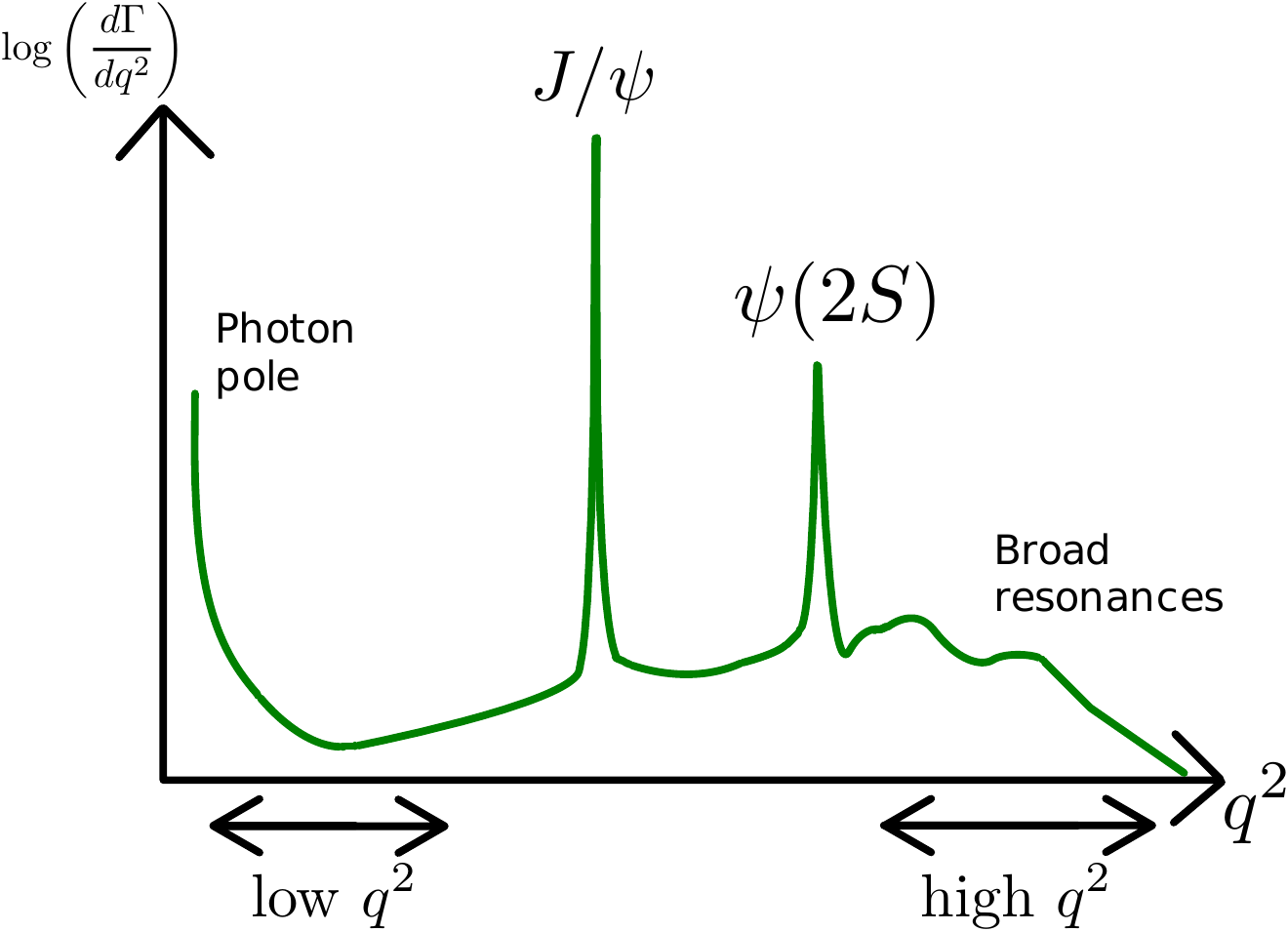}
    \caption{A sketch of the differential width vs.\ \qsq of the decay \decay{\Bdb}{\Kstarzb\mup\mun}. At low \qsq, the effect of diagrams with a virtual photon is clearly seen, while at $\qsq \approx 9.6\gevgevcccc$ and $\qsq \approx 13.6\gevgevcccc$ the amplitude of the charmonium $\cquark\cquarkbar$ states completely dominate. In the region above the \psitwos there is significant interference between broad $\psi$ states and the FCNC amplitude.}
    \label{fig:q2sketch}
\end{figure}

For any final state involving hadrons and a pair of identical charged leptons, there will also be amplitudes from $\bquark \to \cquark\cquarkbar \squark$ where the $\cquark\cquarkbar$ pair subsequently decays electromagnetically to a pair of leptons. The process will have both non-resonant components and resonant components such as the decay $\decay{\Bm}{\Km\jpsi}$ followed by $\decay{\jpsi}{\mup\mun}$. Despite having the same final state, these processes are not FCNC and the amplitudes have to be treated as nuisance effects in searches for NP contributions.
The different \qsq regions are illustrated in Fig.~\ref{fig:q2sketch}: At very low \qsq the decay is dominated by virtual photons coupling to the lepton pair, at low \qsq the penguin amplitude dominates, at intermediate \qsq the narrow \jpsi and $\psi(2S)$ resonances dominate; and at high \qsq the penguin amplitude dominates again but with significant interference from the broad $\psi$ resonances. The regions of the \qsq spectrum with the most precise SM predictions are the low region $\qsq<1\gevgevcccc$ and the high region $\qsq>15\gevgevcccc$.

\subsection{Combinations of multiple measurements}
A measurement of a single electroweak penguin decay is not able to provide a comprehensive picture of any influence of NP. Decays are affected differently by NP and are affected in very different ways by the theoretical uncertainties arising from the form factor calculations and the effect of the \cquark\cquarkbar intermediate states.

As an example, while the $\bquark \to \squark$ processes are the dominant electroweak penguin processes of \bquark-hadron decays, $\bquark \to \dquark$ processes are also important. In the SM, they are Cabibbo suppressed with respect to the $\bquark \to \squark$ processes. However, it is only in what is called Minimal Flavour Violation models~\cite{DAmbrosio:2002vsn} that the same Cabibbo suppression will exist for NP amplitudes. It is thus important to study the rarer $\bquark \to \dquark$ processes as they may be able to give insights to how any NP particles interact with the three generations of quarks. Another example is the study of the decay \decay{\Bsb}{\mup\mun} which has a very small uncertainty in the SM prediction of its branching fraction but on the other hand gives information only on information on the scalar, pseudoscalar and axial vector currents.

\begin{figure}
    \centering
    \includegraphics[width=0.9\linewidth]{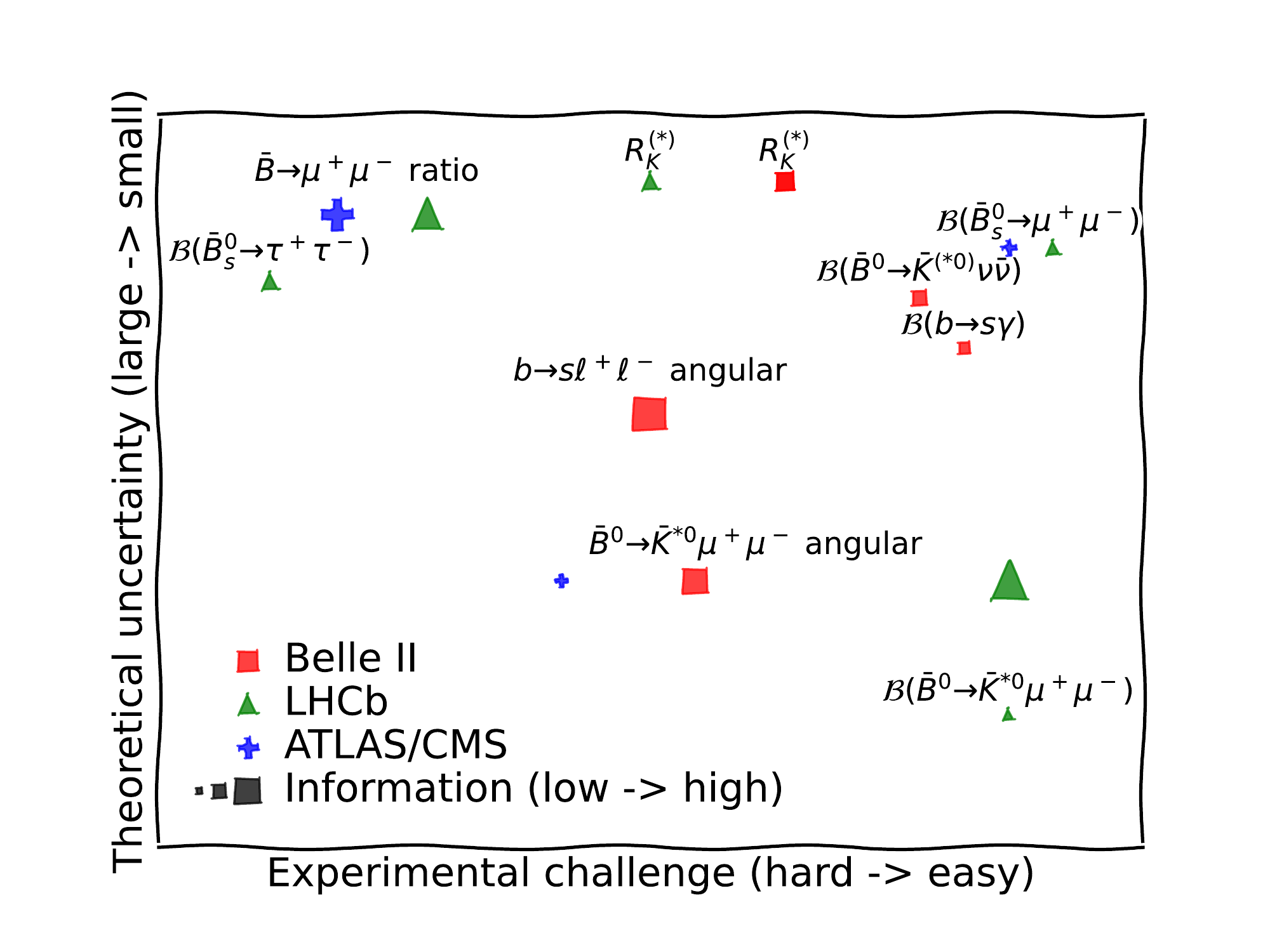}
    \caption{An overview of how different measurements present different experimental challenges and have different accuracy in their SM predictions. The marker size indicates the amount of information that the measurement will provide.}
    \label{fig:overview}
\end{figure}

Figure~\ref{fig:overview} shows an overview of different electroweak penguin measurements. Each type of measurement at a given experiment has been given a subjective rating in terms of how easy it is to perform and what the associated theoretical accuracy is for the SM predictions. The markers in the figure indicate the amount of information that can be gained from the measurement. As an example, the measurement of the ratio of branching fractions between \decay{\Bdb}{\mup\mun} and \decay{\Bsb}{\mup\mun} is experimentally very challenging due to the low branching fraction of \decay{\Bdb}{\mup\mun} but has a very accurate SM prediction. It will thus sit in the top left of the diagram. A deviation from the SM prediction will provide a single number on the coupling structure of NP to different generations and will thus give an intermediate amount of information. Conversely, the \decay{\Bzb}{\Kstarzb\mup\mun} angular analysis is much easier experimentally, has larger theoretical uncertainties and provides a large number of observables. It thus has a large marker towards the bottom right of the diagram.
\begin{marginnote}[]
\entry{Observable}{The expression observable is used for a property that can measured about a given decay. This can be the branching fraction, a lifetime or a measurement related to the angular distribution of the decay products. While a two-body decay will only have a few observables, there can be many for a multibody decay.}
\end{marginnote}

The remainder of this review is divided up into sections that represent the different experimental challenges in measuring the properties of all these decays. Rather than providing a comprehensive overview of all measurements made, examples are given that illustrate different experimental methods. For individual experimental measurements as well as averages over them, the Heavy Flavor Averaging Group~\cite{HFLAV18} and the Particle Data Group~\cite{PDG2020} offer the most comprehensive information.

\section{EXPERIMENTAL CHALLENGE} 

All ground states of \bquark hadrons are characterised by relatively long lifetimes of order of $10^{-12}$ seconds and a large number of accessible final states. Lifetimes are long as the decays are governed by the weak force and the applicable CKM matrix elements \Vcb and \Vub both have a small magnitude. The lifetimes are key-parameters to access physics information (neutral $B$-meson mixing frequencies and CP violation parameters) but also to reject backgrounds. As FCNC decays have branching fractions of $\sim 10^{-5}$ or below, very large datasets have to be recorded to obtain significant signal samples. Another point of paramount importance is the ability to perform particle identification. Electron, muon and photon identification is essential for identifying the decays, while the capability to separate pions, kaons and protons is essential to reduce cross feed between different electroweak penguin decays.

\begin{figure}[tb]
  \centering
 \includegraphics[angle=0,width=0.57\textwidth]{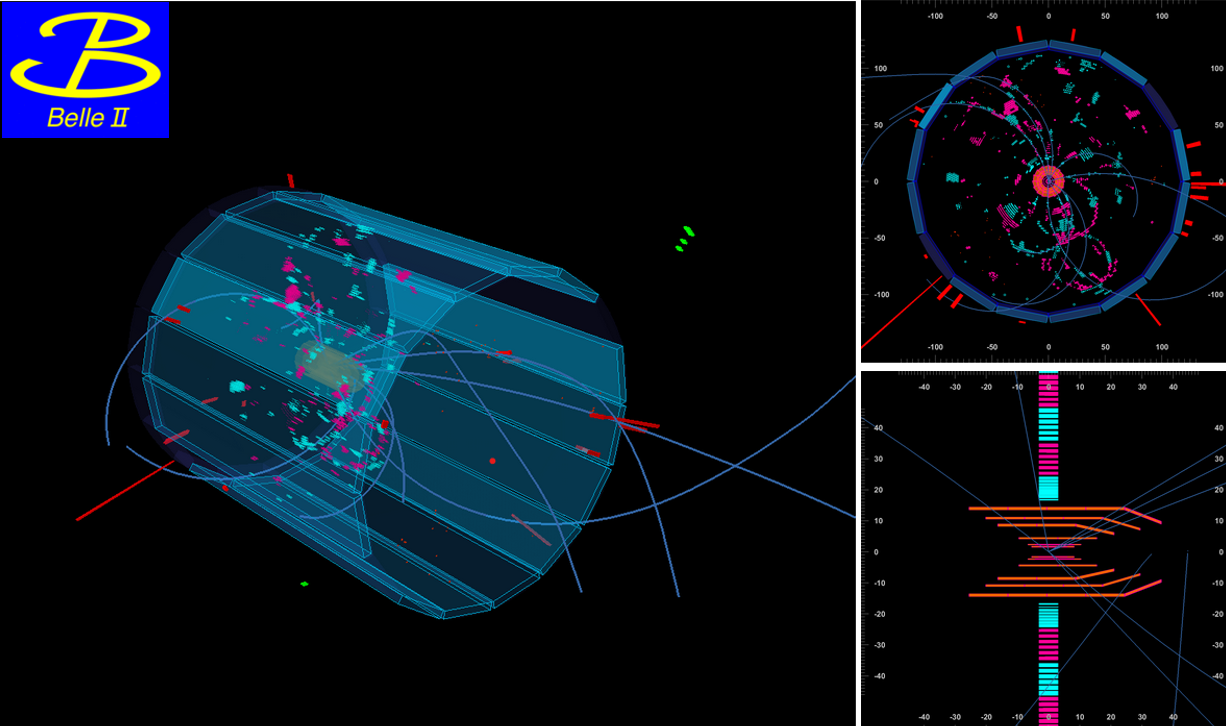} \\
\vspace{-5mm}
 \includegraphics[angle=0,width=0.4\textwidth]{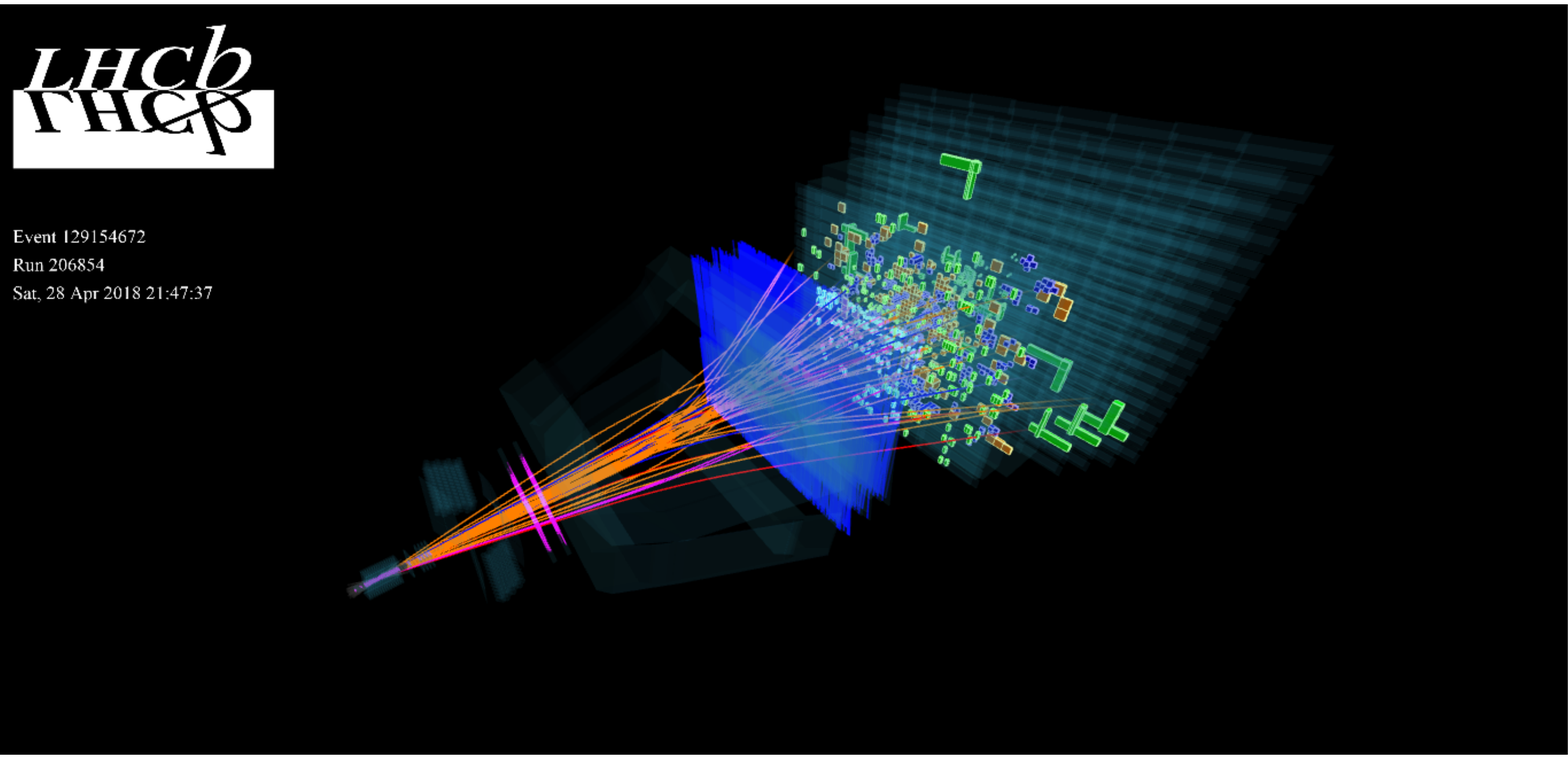}
 \caption{ Event displays of (top) \belletwo\ and (bottom) \lhcb experiments. The very different experimental configurations in terms of solid angle and occupancy in the event are very visible. At \belletwo\ the \epem collisions occur in the centre of the detector, whereas for \lhcb, the proton-proton collisions occur to the far left.} 
\label{fig:EvtDisplay}
\end{figure}

The experimental environments for the experiments that are currently active in the measurement of electroweak penguin decays of \bquark hadrons are very different: the \belletwo experiment at the KEK \B-factory and the \lhcb experiment running at the \lhc proton-proton collider. Event displays, clearly exhibiting the differences, are shown in Fig.~\ref{fig:EvtDisplay} for the two experiments.
\begin{marginnote}[]
\entry{$B$-factory}{An \epem collider running at a centre-of-mass energy corresponding to the mass of the \FourS\ resonance. The $B$-factory name is due to the value of \FourS\ resonance cross section ($\sim 1.1 \nb$) and to the fact it decays uniquely into a pair of $B$-mesons.}
\end{marginnote}

At \B-factories, the \FourS\ resonance produced in the \epem collisions decays into either a $\Bp-\Bm$ pair or a  $\Bz - \Bzb$ pair, and thus represents a copious source of charged and neutral $B$ mesons in a low-background environment. The beam energies at a \B factory are not the same for the \ep and the \en beams meaning that, even if the \B mesons are produced almost at rest in the \FourS\ rest frame, they will fly along the beam axis before they decay and the flight distance can be used to measure their decay time. Since the boost is not very large, the detector is only slightly asymmetric and covers nearly the full solid angle.
\begin{marginnote}[]
\entry{\FourS\ resonance}{The $\Upsilon$ system is the bound state of a \bquark and a \bquarkbar quarks in a $J^{PC}=1^{--}$ state. Various excited states (resonances) exists. The fourth one, called \FourS, has just sufficient energy to create a $\Bp-\Bm$ or a $\Bz-\Bzb$ pair.}
\end{marginnote}

On the contrary, at the LHC the \bquark and \bquarkbar quarks are predominantly produced colinearly, close to the beam direction. This configuration, due the fact that the  main mechanism for \bbbar production is gluon-gluon fusion, also produces uncorrelated \bquark-hadrons species. This geometrical characteristic has strongly influenced the design of the \lhcb detector, which is a single-arm forward spectrometer with a polar-angle coverage between 10 and 300 mrad in the horizontal plane and 250 mrad in the vertical one. 

For the different experiments, the current size of the events samples as well as the \bquark-hadron species available are summarised in 
Tab.~\ref{tab:Exp}. At the beginning of the next decade, the LHCb experiment is planning to have recorded about 50\invfb and the Belle II experiment about 50\invab. The LHCb experiment is proposing a further upgrade that will enable the collection of 300\invfb by the end of the High-Luminosity LHC period~\cite{LHCb-PII-Physics}. 

\begin{table}[htpb]
\centering
\caption{Principal characteristics of the data samples accumulated by the end of 2021 at the $B$-factory experiments (\babar\ (1999-2008), \belle\ (1999-2010) and \belletwo\ (2019-) at the \FourS\ resonance and the \lhcb (2009-) experiment in proton-proton collisions.}
\label{tab:Exp}
\resizebox{\textwidth}{!}{
\renewcommand\arraystretch{1.2}
\begin{threeparttable}
\begin{tabular}{c|c|c|c|c}
{\textbf{Experiment}}	& \textbf{Integrated }	&  $\boldmath{\bbbar}$ \textbf{cross section} &\textbf{Hadronic}  &  \textbf{Main \bquark-hadron species}\\
 & \textbf{ luminosity}& &\textbf{background}  & \textbf{species produced } \\
 \hline
\babar\      & $ 433\invfb$ 	& 1.1 nb & 3.7 nb & \Bzb and \Bm \\
\belle\      & $711\invfb$ 	& 1.1 nb & 3.7 nb & \Bzb and \Bm \\
\belletwo\   & $195 \invfb$ & 1.1 nb & 3.7 nb & \Bzb and \Bm \\
\lhcb\ 		 & $\phantom{00}9\invfb$ & $140 \  \mu$b   & 60 mb  & \Bzb, \Bm, \Bsb, $\Lambda_b$ and  \Bcm\\ 
\atlas\tnote{a}      &  $173\invfb$ & $140 \  \mu$b   & 60 mb  & \Bzb, \Bm, \Bsb, $\Lambda_b$ and  \Bcm \\
\cms\tnote{a}        & $178\invfb$ & $140 \  \mu$b   & 60 mb  & \Bzb, \Bm, \Bsb, $\Lambda_b$ and  \Bcm\\ 
\end{tabular}
\begin{tablenotes}
    \item[a] Only a small fraction of the trigger bandwidth in \atlas\ and \cms\ is dedicated to \bquark hadron physics.
\end{tablenotes}
\end{threeparttable}
}
\end{table}

With the \bquark-hadrons having very different production mechanisms at \B factories and the \lhc, the background levels are very different. At $B$-factories, the non-\bbbar background to signal ratio is of the order of four, while it is several hundred for \lhcb. Consequently, while the trigger for \bquark-hadron events is fully efficient at $B$-factories, it is significantly lower at \lhcb, despite the usage of a complex trigger system consisting of a hardware stage, based on information from the calorimeter and muon systems, followed by a software stage, which applies full event reconstruction. Since the occupancy of the calorimeters is significantly higher than that of the muon stations, the constraints on the trigger rate require that higher thresholds are imposed on the electron transverse energy than on the muon transverse momentum. The most noticeable effect on analyses testing Lepton Universality described in this review is that while at $B$-factories the yields of \bTosmm and \bTosee transitions are similar, a factor of 4 to 5 in favour of the muonic modes is observed at \lhcb. From 2022 with the upgrade of the \lhcb experiment, and in particular the removal of the hardware trigger, accessing higher yields for the modes with electrons may be possible. 

In order to distinguish signal from background events, both the \belletwo and \lhcb experiments exploit the reconstruction of the \bquark-hadron invariant mass from the measured decay products. At \belletwo, the background is reduced thanks to the constraints provided by the \FourS\ resonance. Since it decays uniquely into a pair of $B$ mesons, the energy of the $B$-decay products is equal to half of the centre-of mass energy. This enables two weakly correlated discriminating variables to be defined: one which compares the reconstructed $B$-meson energy to the beam energy and one which corresponds to the $B$-meson mass reconstructed from the measured momenta of the decay products and the beam energy. While the proton-proton collision environment for \lhcb doesn't provide a beam energy constraint, the boost of the \bquark-hadrons results in decay lengths of the order of 1\cm. The reconstruction of secondary vertices from the \bquark-hadron decay leads to a very large rejection of tracks originating from the primary collision vertex.
\begin{marginnote}[]
\entry{Secondary vertex}{The point where the \bquark-hadron decays at a hadron collider. The event will also have one or more primary vertices where the proton-proton collision happens.}
\end{marginnote}

\begin{figure}[htpb]
\centering
 \includegraphics[angle=0,width=0.3\textwidth]{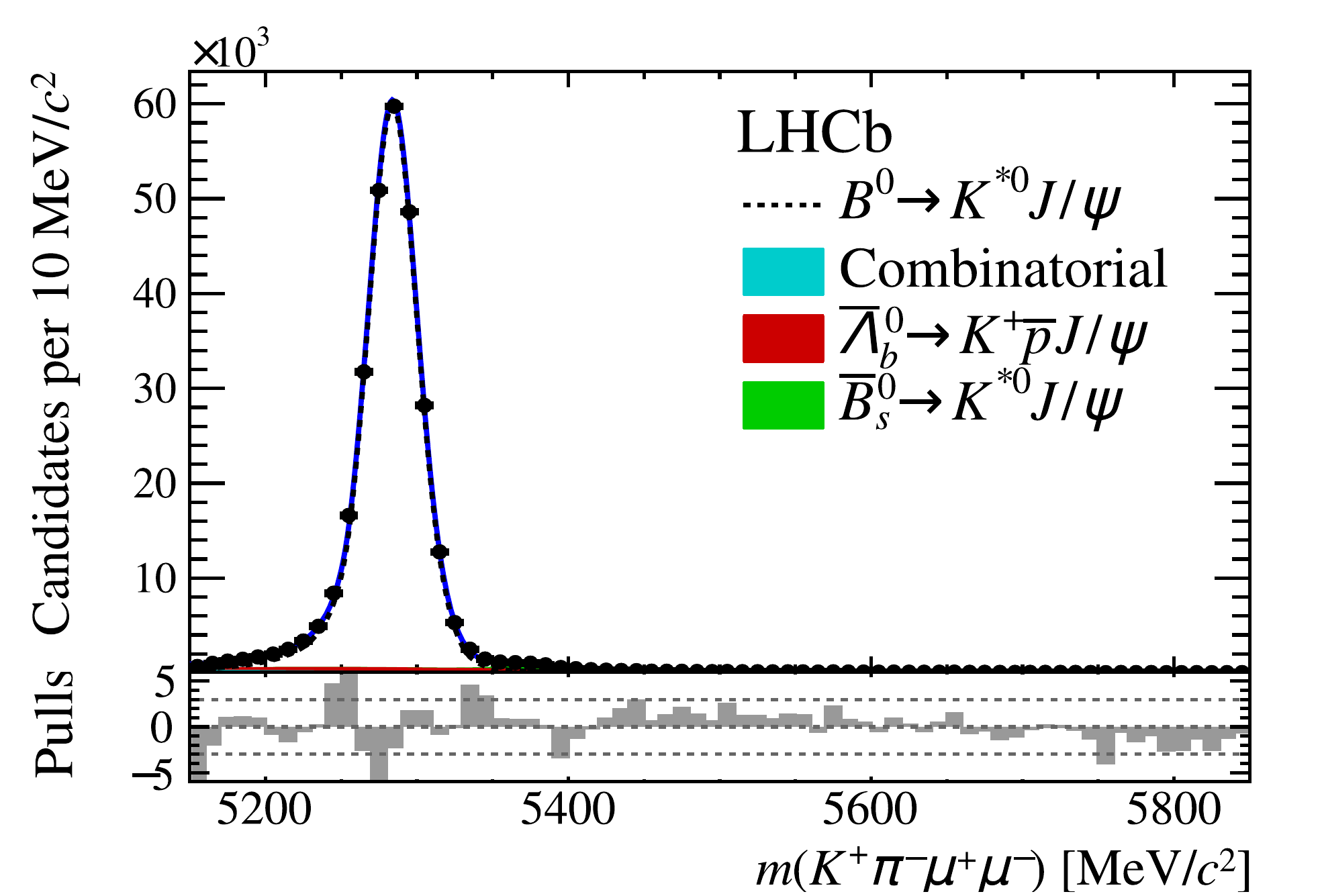}
 \hskip 2.5cm
  \includegraphics[angle=0,width=0.3\textwidth]{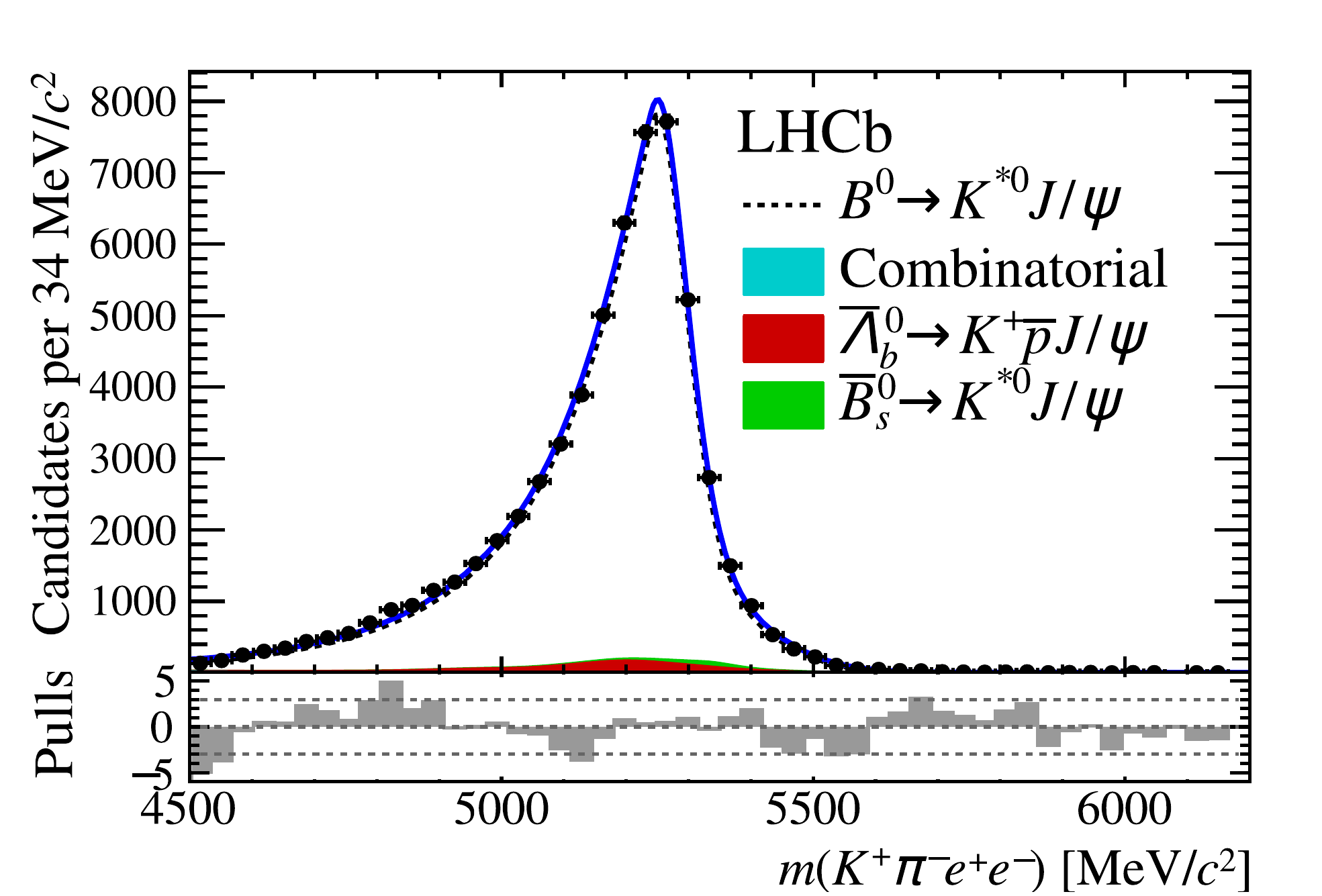}
  \vskip .5cm
   \includegraphics[angle=0,width=0.3\textwidth]{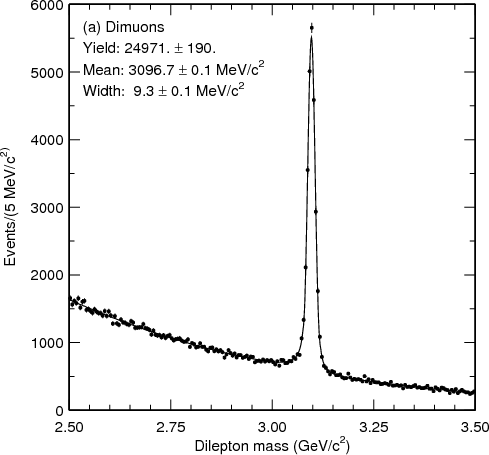}
  \hskip 2.5cm
    \includegraphics[angle=0,width=0.3\textwidth]{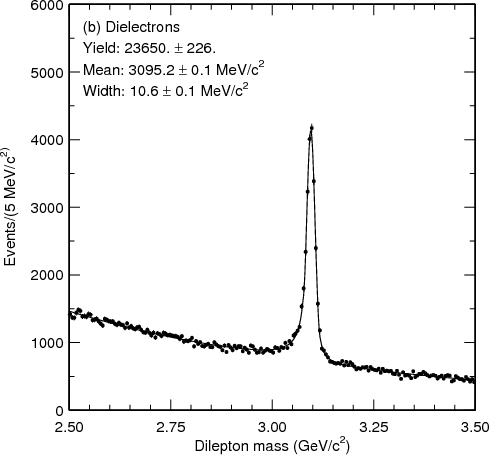}

\caption{Top: distribution of the four-body invariant mass in \BdToKstKpiJPsll decays for (left) $\ell=\mu$ and (right) $\ell=e$ reconstructed at \lhcb. From Ref.~\cite{LHCb-PAPER-2017-013}.  Bottom: distribution of the dilepton invariant mass in the \jpsi region for (left) $\ell=\mu$ and (right) $\ell=e$ reconstructed at \belle. From Ref.~\cite{Abe:2002rc}.}
\label{fig:ExpBrem}
\end{figure}

Another experimental difference, of particular importance for this review, is the difference in the amount of bremsstrahlung emission due to the difference in energies and detector designs. Dedicated recovery procedures are in place in order to improve momentum reconstruction. Despite these corrections, the \B mass resolution is still degraded for final states involving electrons compared to final states including muons. While the effect is subtle at \B-factories, it is quite significant for \lhcb as can be seen from Fig.~\ref{fig:ExpBrem}. 

In order to reduce the systematic uncertainties at the lowest possible level, control modes are of prime importance for many aspects of the analyses described in this review. Such modes are used in particular to correct the simulation to allow for the extraction of the efficiencies of the signal selection with respect to particle identification, track reconstruction, trigger and \bquark-hadron kinematics. Since simulating precisely the \bquark-hadron production characteristics as well as the subsequent detector response in the busy hadronic environment of the \lhc is challenging, the use of control modes is a key feature of many of the \lhcb analyses presented here.

In addition to the \B-factories and the \lhcb experiment, the \atlas and \cms experiments at the \lhc have also performed analyses of electroweak penguin decays. However, the limited trigger bandwidth devoted to \bquark hadron decays and the lack of hadron particle identification means that competitive measurements have only been made in final states with a pair of muons.

Most of the analyses presented in this review use Machine Learning techniques to distinguish signal from background, to ensure the largest possible signal yields. Since the background rejection is already a key issue at the trigger level for LHCb, such techniques are also implemented at the software trigger level~\cite{LHCb:2018zdd}. 

\section{BRANCHING FRACTION MEASUREMENTS} 
The measurement of a branching fraction that is integrated over the complete phase space of the final state is conceptually an easy measurement. The final state can be either exclusive (e.g.~\decay{\Bsb}{\mup\mun}), a sum over exclusive states (e.g.~\decay{\Bzb}{\Kstarzb\neu\neub}, where all neutrino types are included) or inclusive (e.g.~\decay{\bquark}{\squark \gamma}, where there is a sum over all final states containing net strangeness.). 
\begin{marginnote}[]
\entry{Exclusive vs. inclusive}{An exclusive measurement is with a specific initial and final state, like \decay{\Bdb}{\Kstarzb\mup\mun} while an inclusive measurement sums over all decays included in a given quark level transition like \decay{b}{s\mup\mun}.} 
\end{marginnote}

At hadron colliders, branching fraction measurements are performed by making a measurement relative to a another branching fraction that has been measured in the past. These type of normalisation measurements are performed at \B-factories where the total number of \FourS\ produced can be determined without looking at specific decay modes. For \Bz and \Bp decays, this type of normalisation works well, but for \Bs, \Lb and \Bc decays this works less well. As these particles are not produced in \FourS\ decays there are no branching fractions that can act as normalisation modes. Instead \Bz or \Bp modes are used, which is turn requires that the relative production fractions of \eg\ \Bs and \Bp is known. These production fractions are measured by comparing the rate of specific \Bs and \Bp decays that can be related through SU(3)~\cite{LHCb:2021qbv}. For several branching fraction measurements of electroweak penguin decays, the uncertainty on the overall normalisation procedure is a limiting systematic. 

\subsection{Leptonic decays}
The SM prediction for the branching fraction of the \decay{\Bsb}{\mup\mun} decay is $(3.66\pm0.14)\times 10^{-9}$ and for the Cabibbo suppressed \decay{\Bdb}{\mup\mun} decay $(1.03\pm0.05)\times 10^{-10}$~\cite{Bobeth:2013uxa, Beneke:2019slt}. As the initial state is a spin zero pseudoscalar and the final states contain only the two leptons, the left-handed nature of the weak interaction forces one of the muons to be in the wrong helicity state making the decay \textit{helicity suppressed}. On the other hand, with no hadrons in the final state, the QCD uncertainty is small. For SM predictions this results in tiny branching fractions with a very small relative uncertainty. Gaining information from these decays is thus completely limited by experimental issues.

\begin{figure}
    \centering
    \includegraphics[width=0.75\linewidth]{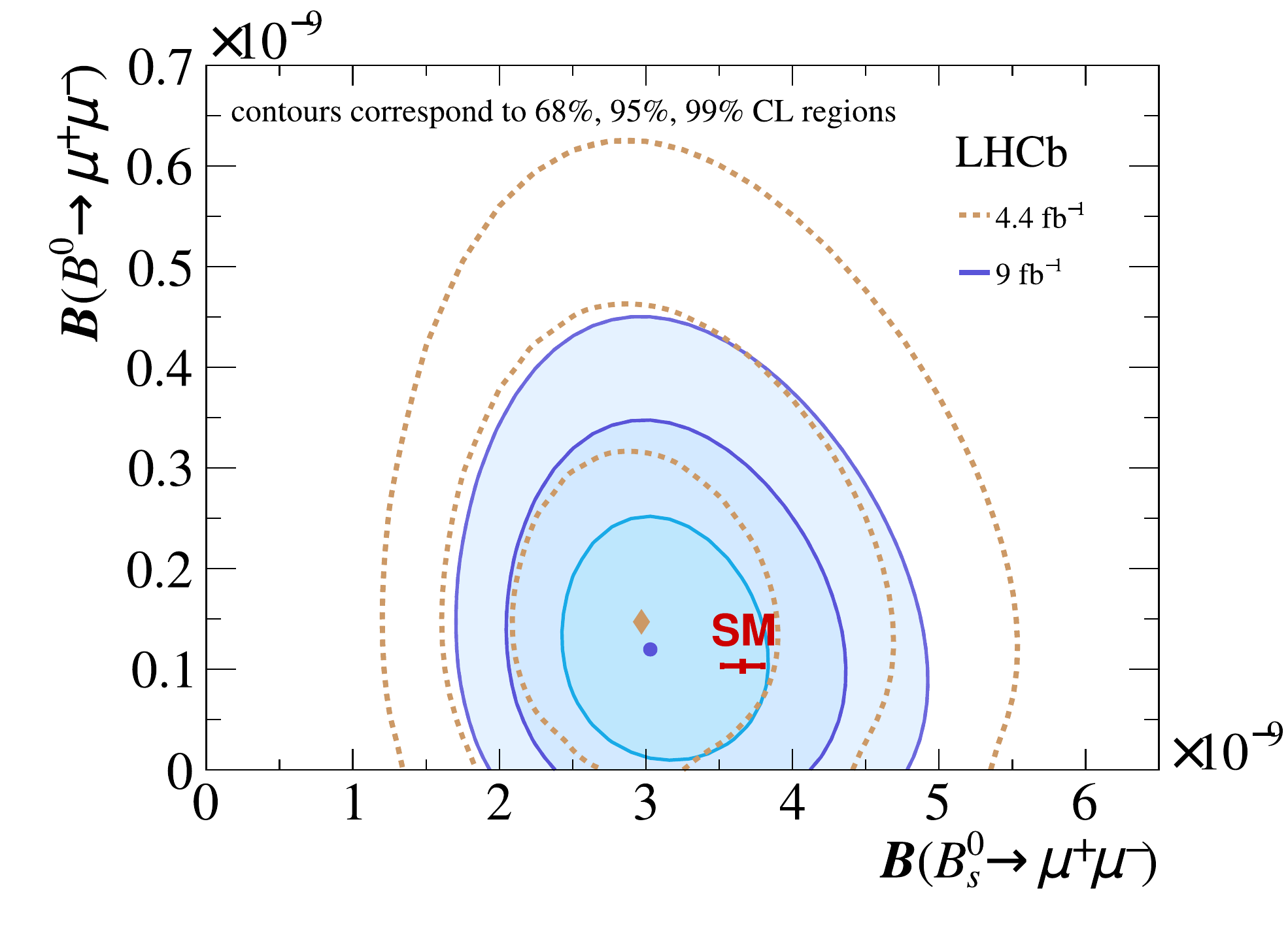}
    \vspace{-0.5cm}
    \caption{Contour lines for the confidence limits on the branching fractions of \decay{\Bzb}{\mup\mun} versus \decay{\Bsb}{\mup\mun}. From Ref.~\cite{LHCb:2021awg}.}
    \label{fig:B0VsBsMuMu}
\end{figure}

With just the two muons in the final state, the signature of the decays is easy to trigger on and to reconstruct in a hadronic environment, allowing measurements to be made by the three LHC experiments \atlas~\cite{ATLAS:2018cur}, \cms~\cite{CMS:2019bbr} and \lhcb~\cite{LHCb:2021vsc}. The trigger selects initially either one or two muons with a large transverse momentum with respect to the beam axis. In the subsequent selection, the signature is two muons with a vertex displaced from the primary proton-proton interaction vertices. Combined with the very low misidentification of pions as muons, the final state has very low background and the main challenges are to have a low trigger threshold and to collect a large amount of integrated luminosity to see the rare decays. The observation of the \decay{\Bsb}{\mup\mun} decay was made in a combined analysis by the \cms and \lhcb collaborations~\cite{LHCb-PAPER-2014-049} and the current most precise measurement is $\mathcal{B}(\decay{\Bsb}{\mup\mun}) = (3.09\PMerr{0.46}{0.43}\PMerr{0.15}{0.13})\times 10^{-9}$~\cite{LHCb:2021vsc}. Here the first uncertainty is the statistical uncertainty, while the second is the systematic. The systematic uncertainty is dominated by the normalisation of the branching fraction measurement, that is performed relative to the \decay{\Bm}{\jpsi\Km} decay. 
There is still only a limit on the \decay{\Bdb}{\mup\mun} decay but if the \decay{\Bdb}{\mup\mun} decay has the SM branching fraction, an observation is expected during Run 3 of the LHC (2022-2025) while the decay will remain out of reach for \belletwo for the foreseeable future due to the extremely low branching fraction. As can be seen from Fig.~\ref{fig:B0VsBsMuMu}, current measurements are in good agreement with the Standard Model.

While it might appear that the decay \decay{\Bdb}{\neu\neub} should be able to proceed through a box diagram with two \W bosons, it is in fact forbidden in the SM as it would require a right-handed neutrino or left-handed antineutrino. Searching for this decay therefore turns into a search for the $B$ meson to decay into anything invisible.

\subsection{Radiative decays}

Radiative decays of $b$-flavour hadrons proceed through
\btosgam or \btodgam penguin diagram.
In 1993, CLEO reported the first observation of the $\Bb \to \Kstarb\gamma$
decay~\cite{Ammar:1993sh}, which was the first experimental
confirmation of a penguin decay.
The branching fraction of $\Bb \to \Kstarb\gamma$ has subsequently been measured by other experiments, and the current world average gives
$\mathcal{B}(\Bb \to \Kstarb\gamma) = (4.18 \pm 0.25) \times 10^{-5}$~\cite{PDG2020}.
Many other exclusive $\Bb$ decays such as $\Bb \to \Kbar_1(1270)\gamma$,
 $\Bb \to \Kbar_2^*(1270)\gamma$, $\Bb \to \Kbar\eta\gamma$ have also been measured.
Radiative decays of $\Bsb$ mesons have been studied with \lhcb
as well as Belle data taken at the $\FiveS$ resonance, and
the branching fractions have been measured.
The radiative $b$-baryon decay $\Lambda_b \to \Lambda\gamma$ has been
observed by LHCb~\cite{Aaij:2019hhx}.

\begin{marginnote}[]
\entry{\FiveS\ resonance}{The fifth $\Upsilon$ resonance $\FiveS$ (or $\Upsilon(10860)$) can decay to $\Bs^{(*)}\Bsb^{(*)}$. Belle took 121\invfb of data mainly for the study of \Bs decays.}
\end{marginnote}

Measurement of the branching fractions of the exclusive decays is
experimentally straightforward. On the other hand, the prediction of the exclusive branching fractions suffers from large uncertainties due to uncertainties in the hadronic form factors, and the comparison with experimental results does not provide much information. In terms of the theoretical calculation, the present SM prediction with NNLO calculation is
$\mathcal{B}(\btosgam) = (3.31 \pm 0.23) \times 10^{-4}$~\cite{Misiak:2015xwa}
for $E_\gamma > 1.6~\gev$~\cite{Misiak:2015xwa}.
This is much more precise than the prediction for exclusive processes,
and therefore one can compare it with the experimental measurements
in order to search for or constrain NP.

Measurement of the inclusive \btosgam
(denoted by $B \to X_s\gamma$, where $X_s$ is a hadronic system including
$s$ quark)
branching fraction
is challenging and is only possible in a clean environment of  \B factories. 
One approach is to measure the $\gamma$ spectrum in
$e^+e^- \to \Upsilon(4S) \to B\Bb$.
The raw photon spectrum includes a huge background from
photons originating from the continuum $e^+e^- \to \qqbar$,
and this contribution is subtracted using data taken at a collision energy
slightly below that at which the $\Upsilon(4S)$ is produced.
There still exists large background from $B$ decays that do not
come from \btosgam, and this contribution is estimated by simulation.
A high momentum lepton which arises from semi-leptonic decays of the
other $B$ meson (lepton tag)
is often required to suppress the continuum background.

\begin{figure}
 \centering
 \includegraphics[angle=0,width=0.38\textwidth]{%
 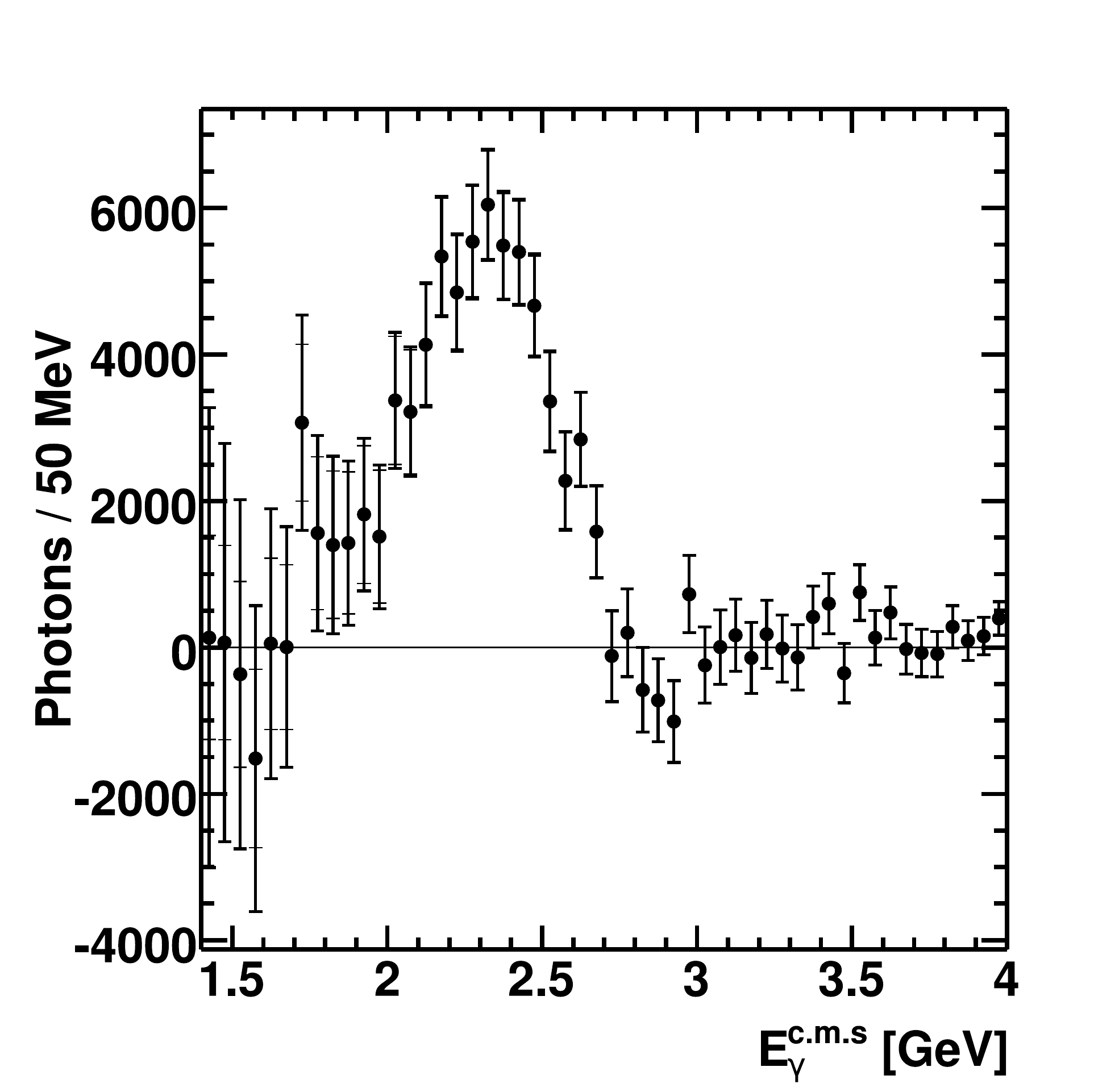}
 \caption{\btosgam spectrum,
 without the correction of the signal acceptance,
 obtained by Belle~\cite{Limosani:2009qg}.
 The lepton tag is not applied to obtain this distribution.
 }
 \label{fig:bsgamma-belle}
\end{figure}

Figure~\ref{fig:bsgamma-belle} shows the photon energy spectrum
in $B \to X_s\gamma$ obtained by Belle
with $605~\invfb$ data~\cite{Limosani:2009qg}.
Strictly speaking, this includes the contribution of \btodgam described
later.
As seen in the figure, the photon spectrum peaks around $2~\gev$.
The measurement of energy bins below $2.0~\gev$ is difficult
because of a small signal and a large increase of background from other $B$ decays.
For this reason, the inclusive \btosgam branching fraction is quoted
with a certain minimum energy.
The photon energy spectrum includes information
of the mass and dynamics of the \bquark quarks in the \B meson,
which are useful in the estimation of $|\Vub|$ and $|\Vcb|$
from the measurements of semi-leptonic \B decays~\cite{Bosch:2004th}.

It is also possible to suppress the background further by
fully reconstructing one of the \B mesons
from $e^+e^- \to \FourS \to \B\Bb$ and then measure the photon
spectrum~\cite{BaBar:2007yhb}. This full reconstruction method gives a signal with much higher purity at the cost of reduced statistics due to an low efficiency of less than $1\%$. This is a promising method for the future when \belletwo accumulates
an order of magnitude higher data sample.

Another approach in the measurement of $\B \to X_s\gamma$
is to reconstruct the $X_s$ part from many final states such as $K\pi$ and $K\pi\pi$.
This method has higher purity than the normal inclusive analysis
and has the advantage that it is possible to distinguish between \B and \Bb,
and to separate $X_d$ from $X_s$.
However, one needs to rely on simulation for the hadronisation of $X_s$,
and the contribution of non-reconstructed $X_s$ states
gives rise to large uncertainty.

The present average of the measurements is
$\mathcal{B}(\btosgam) = (3.27 \pm 0.15) \times 10^{-4}$
at $E_\gamma > 1.6~\gev$, which agrees well with the SM prediction.
This measurement sets a strong constraint on the Wilson coefficient $C_7$.

The \btodgam process is suppressed relative to \btosgam by a factor of
$(|\Vtd|/|\Vts|)^2 \sim 0.04$, and the branching fractions
of the exclusive decays are below $10^{-6}$.
Therefore, the analysis of \btodgam needs to cope with larger backgrounds,
compared to the \btosgam analysis.
In addition, the \btosgam process itself becomes a background.
For example, the $\Bb \to \rho^0\gamma, \rho^0 \to \pi^+\pi^-$
signals can be easily contaminated by
$\Bb \to \Kstarzb\gamma$ with \decay{\Kstarzb}{\Km\pip} when a kaon is misidentified as a pion. Good particle identification helps to separate the signal from this type of background.

The measurement of the ratio of $\Bb \to \rho\gamma$ and $\omega\gamma$
branching fractions to that for $\Bb \to \Kstarb\gamma$
allows for the determination of $|\Vtd|/|\Vts|$
with some residual uncertainty due to the ratio of their respective form factors. With the first observation of these decays by Belle~\cite{Belle:2005grh},
$|\Vtd|/|\Vts|$ was measured for the first time to be
$0.199^{+0.026}_{-0.025}{}^{+0.018}_{-0.015}$, where the first and second
errors are the experimental and theoretical errors, respectively.
After the observation of \Bs oscillations by CDF~\cite{CDF:2006imy},
$|\Vtd|/|\Vts|$ is measured with much higher precision from
those, and \btodgam is not used in the determination of 
$|\Vtd|/|\Vts|$.
However, it is still important to measure $|\Vts|/|\Vtd|$ with the \btodgam
process since NP may contribute in a different way to the
different diagrams.

\subsection{Semileptonic decays}
Semileptonic electroweak penguin decays are decays such as $\decay{\Bm}{\Km\mup\mun}$, as illustrated in Fig.~\ref{fig:FCNC}(right), and $\decay{\Bzb}{\Kstarzb\neu\neub}$. They should not be confused with the more common semileptonic decays such as $\decay{\Bm}{\Dstarp\mun\neumb}$, which are dominated by tree level diagrams.

\begin{figure}
    \centering
    \includegraphics[width=0.75\linewidth]{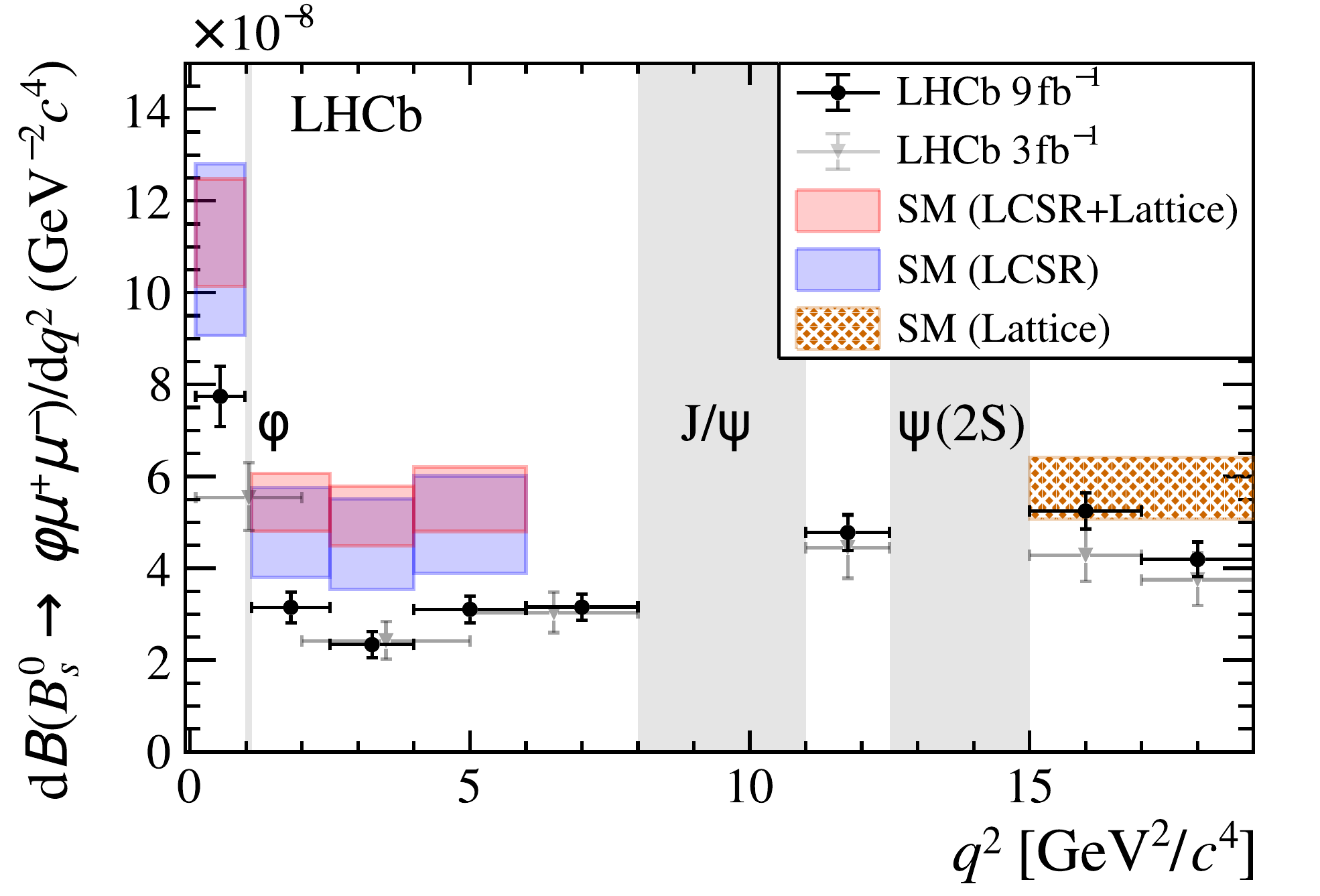}
    \caption{The differential branching fraction of the decay \decay{\Bsb}{\phi\mup\mun}. It can be seen that outside the narrow charmonium resonance regions, marked by the grey vertical bands, the branching fraction is consistently below the SM prediction. From Ref.~\cite{Aaij:2021pkz}}
    \label{fig:BF_BsToPhiMuMu}
\end{figure}

For decays of the type \bsll, measurements of the branching fraction are performed in bins of \qsq, the mass squared of the lepton pair. An example of such a measurement can be seen for the decay $\decay{\Bsb}{\phi\mup\mun}$~\cite{Aaij:2021pkz} in Fig.~\ref{fig:BF_BsToPhiMuMu}. When looking at measurements with non-uniform binning in \qsq, care should be taken to understand if measurements are given as the branching fraction within the bin, or given as a differential branching fraction $\frac{d\B}{d\qsq}$; the latter is obtained from the former by dividing by the bin width. The regions of \qsq that include the narrow \jpsi and \psitwos resonances are excluded from the binned measurements as they are completely dominated by the tree level amplitudes.

The total branching fraction when integrating over \qsq for different decays is obtained by taking the binned measurement and then interpolating them into the missing regions of the \jpsi and \psitwos resonances using a decay model where the resonances are not present. In Fig.~\ref{fig:BF_BsToPhiMuMu}, this amounts to interpolating through the grey vertical bands, ignoring the huge peaks from the resonances. There will still be a level of contamination of this \emph{penguin-only} branching fraction from the broad charmonium resonances present at $\qsq>15\gevgevcccc$. The only way to avoid this is to perform a completely unbinned analysis where all amplitudes in the \qsq spectrum are modelled. This has so far only been done for the \decay{\Bm}{\Km\mup\mun} decay mode~\cite{LHCb:2016due}.

The most precise measurements of branching fractions have been performed for the decays \decay{\Bm}{\Km\mup\mun}~\cite{LHCb:2014cxe}, \decay{\Bzb}{\Kstarzb\mup\mun}~\cite{LHCb:2014cxe} and \decay{\Bsb}{\phi\mup\mun}~\cite{Aaij:2021pkz}. All these branching fraction measurements are currently below the expectations from the SM but with quite significant (and correlated) uncertainties in these SM predictions.
See Sec.~\ref{sec:LFU} on lepton flavour universality for measurements with electrons where the branching fraction measurements currently have most relevance. The leading systematic uncertainty in the branching fraction measurements arises from the normalisation. As an example, the \decay{\Bm}{\Km\mup\mun} branching fraction is measured relative to the decay \decay{\Bm}{\jpsi\Km}, with \decay{\jpsi}{\mup\mun}. The uncertainty on the branching fraction of this normalisation mode is around 5\% and a factor 3 larger than the statistical uncertainty. As \lhcb is unable to provide any precision measurements of branching fractions without a normalisation mode, improvements beyond the 5\% level will only appear once the \belletwo collaboration improves the measurement of the corresponding normalisation mode.

Branching fraction measurements of the decay modes $\decay{\Bb}{\Kbar^{(*)}\neu\neub}$ are very challenging from an experimental point of view, as the only visible final state particle is the kaon. A measurement is not possible at a hadron collider, but at a \B-factory it is possible by exploiting the fact that the \B mesons are produced as a pair from the $\Upsilon(4S)$ decay; once one \B mesons is reconstructed, the kinematics of the other \B meson is known and the only other particles in the detector are the decay products of the \B meson. The current best limit is $\decay{\Bb}{\Kbar\neu\neub} < 3.2 \times 10^{-5}$~\cite{BaBar:2013npw} from the \babar experiment but the analysis method developed by \belletwo in Ref.~\cite{Belle-II:2021rof} sets a very competitive limit of $\decay{\Bm}{\Km\neu\neub} < 4.1 \times 10^{-5}$ with a factor 8 lower integrated luminosity. Rather than relying on an explicit reconstruction of the other \B meson in the event, candidates are selected by identifying charged kaons and then using a machine learning algorithm to test whether the remainder of the event is compatible with the decay products of a single \B meson decay. As the SM prediction for the branching fraction summed over the three neutrino species is $(8.2\pm1.0)\times 10^{-6}$~\cite{Bause:2021ply} an observation is expected soon with data from \belletwo.

\begin{figure}
    \centering
    \includegraphics[width=0.7\linewidth]{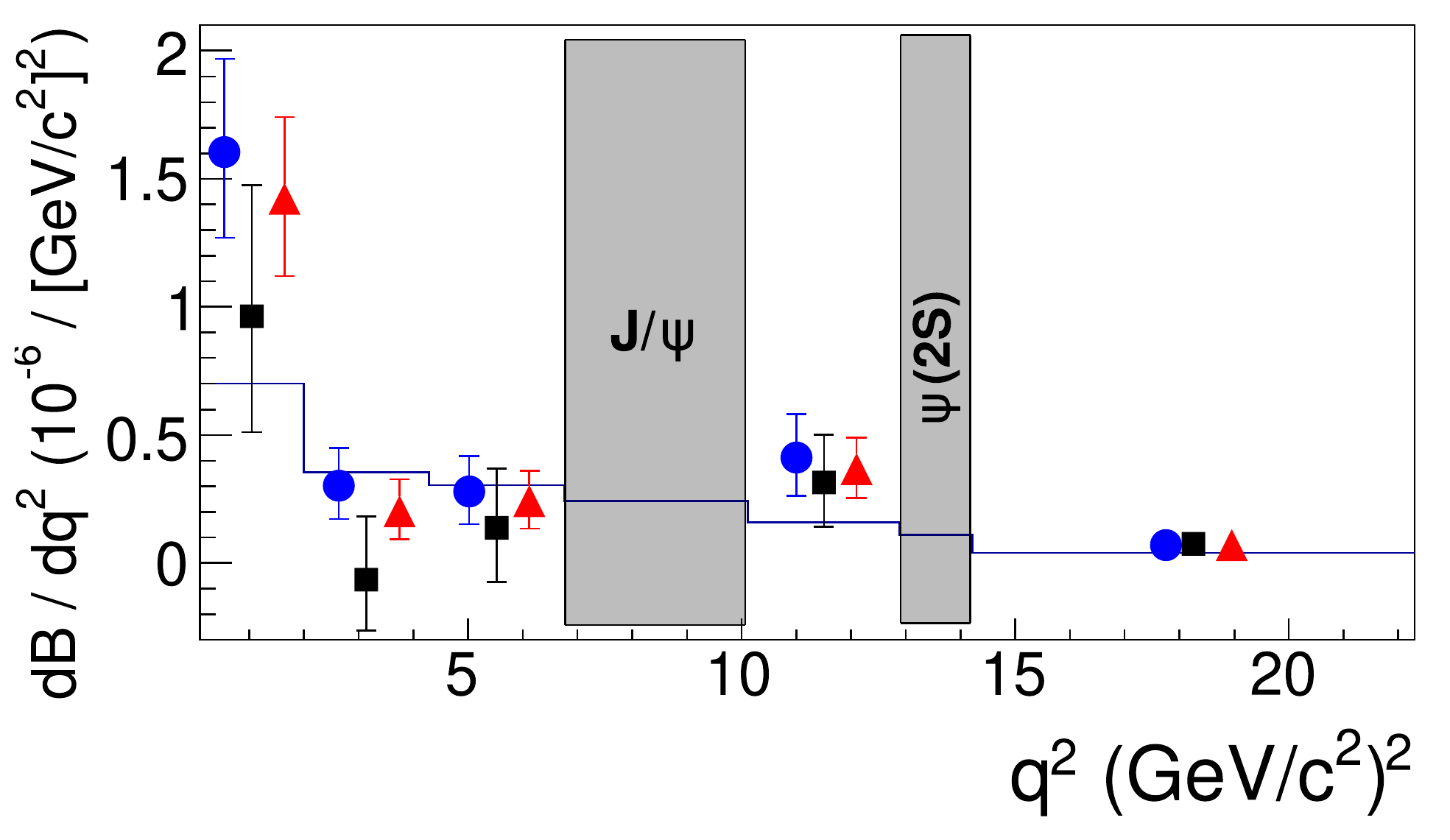}
    \caption{Differential branching fraction for the inclusive $\decay{\B}{X_s\ellell}$ measurement for the final states with (blue) electrons, (red) muons and (black) the lepton flavour average. From Ref.~\cite{BaBar:2013qry}.}
    \label{fig:babar-inclusive}
\end{figure}

In all the decays discussed above the leading uncertainty of around 30\% in the SM prediction is from the $\decay{B}{K}$ form factors as illustrated with the wide bands for the SM prediction in Fig.~\ref{fig:BF_BsToPhiMuMu}. This uncertainty can be reduced to around 7\% with measurements of the inclusive branching fraction of \decay{\B}{X_s\ellell}, where the \B meson represents either a \Bz or a \Bm~\cite{Huber:2020vup}. At \belle~\cite{Belle:2005fli} and \babar~\cite{BaBar:2013qry} this was carried out using a \emph{sum of exclusive modes} approach. In the \babar analysis, a total of twenty final states were used, with either electrons or muons as the lepton pair and the hadronic system containing a neutral or charged kaon and up to four pions. To reduce background, it is required that the mass of the hadronic part of the decay is below 1.8\gevcc. The reconstruction covers around 70\% of the total inclusive rate and simulation is then used to extrapolate to the measurement of the full branching fraction $\mathcal{B}(\decay{\B}{X_s\ellell})=(6.7\pm0.7\pm0.5) \times 10^{-6}$, where the first uncertainty is from the experimental measurement and the second from the extrapolation.
The extrapolated differential rate as a function of \qsq can be seen in Fig.~\ref{fig:babar-inclusive}.
Ideas for carrying out an inclusive analysis at hadron colliders
by identifying decays with a charged kaon and a pair of muons has been suggested in Ref.~\cite{Amhis:2021oik}. The method would extrapolate from the measured rate to the full inclusive rate using isospin assumptions. 

The $\decay{\bquark}{\dquark}$ transition can be measured in the same way as for radiative decays. The most precise measurement is $\mathcal{B}(\decay{\Bm}{\pim\mup\mun})=(1.8\pm0.2) \times 10^{-8}$~\cite{LHCb:2015hsa} which is compatible with the prediction from $|\Vtd|/|\Vts|$ when comparing to the $\decay{\Bm}{\Km\mup\mun}$ decay.

\section{ANGULAR DISTRIBUTIONS}
\label{sec:angular}
The dynamics of \decay{\bquark}{\squark \ellell} decays can be described by helicity angles and the dilepton invariant mass \qsq~\cite{Egede:2008uy, LHCb:2013zuf}. When reading the literature, great care has to be taken in understanding the sign conventions used for the helicity angles and the derived observables as no uniform choice is used. At the LHCb experiment, the most precisely measured decay is \decay{\Bdb}{\Kstarzb\mup\mun}, with \decay{\Kstarzb}{\Km\pip}~\cite{LHCb:2020lmf}. In this case, the dynamics are described by three angles and \qsq. Measurements of such decays have also been made by the ATLAS~\cite{Aaboud:2018krd}, BaBar~\cite{Aubert:2006vb}, Belle~\cite{Wehle:2016yoi}, CDF~\cite{Aaltonen:2011ja} and CMS~\cite{Sirunyan:2017dhj} collaborations. The differential decay distribution is predicted in terms of angular observables~\cite{Ali:1999mm, Kruger:2005ep, Descotes-Genon:2012isb}, and is complicated by the presence of both P~(spin 1) and S-wave~(spin 0) contributions to the \Km\pip system. In the region around the \Kstarzb resonance, the latter is at the $\approx10\%$ level. The contribution from D-wave~(spin 2) \Km\pip is negligible compared to the precision of the present generation of experiments. Correct experimental identification of the kaon and pion is a significant factor in the angular analyses, as the angular convention depends on the charges of the hadrons. 

Measurement of angular observables requires that the data be corrected for the inefficiencies introduced by the effects of the detector, reconstruction and selection as a function of the helicity angles and \qsq. At LHCb, the \emph{acceptance function} that models such effects is determined using simulated events. The simulation is calibrated by comparing it with data in a range of control decays. For example, the particle identification performance of the \lhcb detector is determined with \decay{\Dstarp}{\Dz(\to \Km\pip)\pip}, where the particles can be unambiguously identified on the basis of the kinematics alone; and the \Bd momentum spectrum is calibrated with other fully reconstructed \bquark hadron decays. Calibration of the momentum spectrum is required, as efficiencies depend on quantities in the lab frame, while they need to be applied to the angular distribution in the \Bd rest frame. The overall quality of the calibration is cross-checked with an angular analysis of  \decay{\Bdb}{\jpsi(\to \mup\mun)\Kstarzb(\to \Km\pip)} decays. Such decays are mediated by a tree-level process and hence occur $\sim100$ times more often than the \decay{\Bdb}{\Kstarzb\mup\mun} rare decay (see Fig.~\ref{fig:q2sketch}). The angular observables are known precisely from previous measurements and can therefore be used to validate the simulation description of the acceptance function. 
\begin{marginnote}[]
\entry{Angular observable}{A combination of the helicity amplitudes which can be measured experimentally and which is developed such that it minimises theoretical uncertainty from form factor calculations.} 
\end{marginnote}

\begin{figure}
    \centering
    \includegraphics[width=0.75\linewidth]{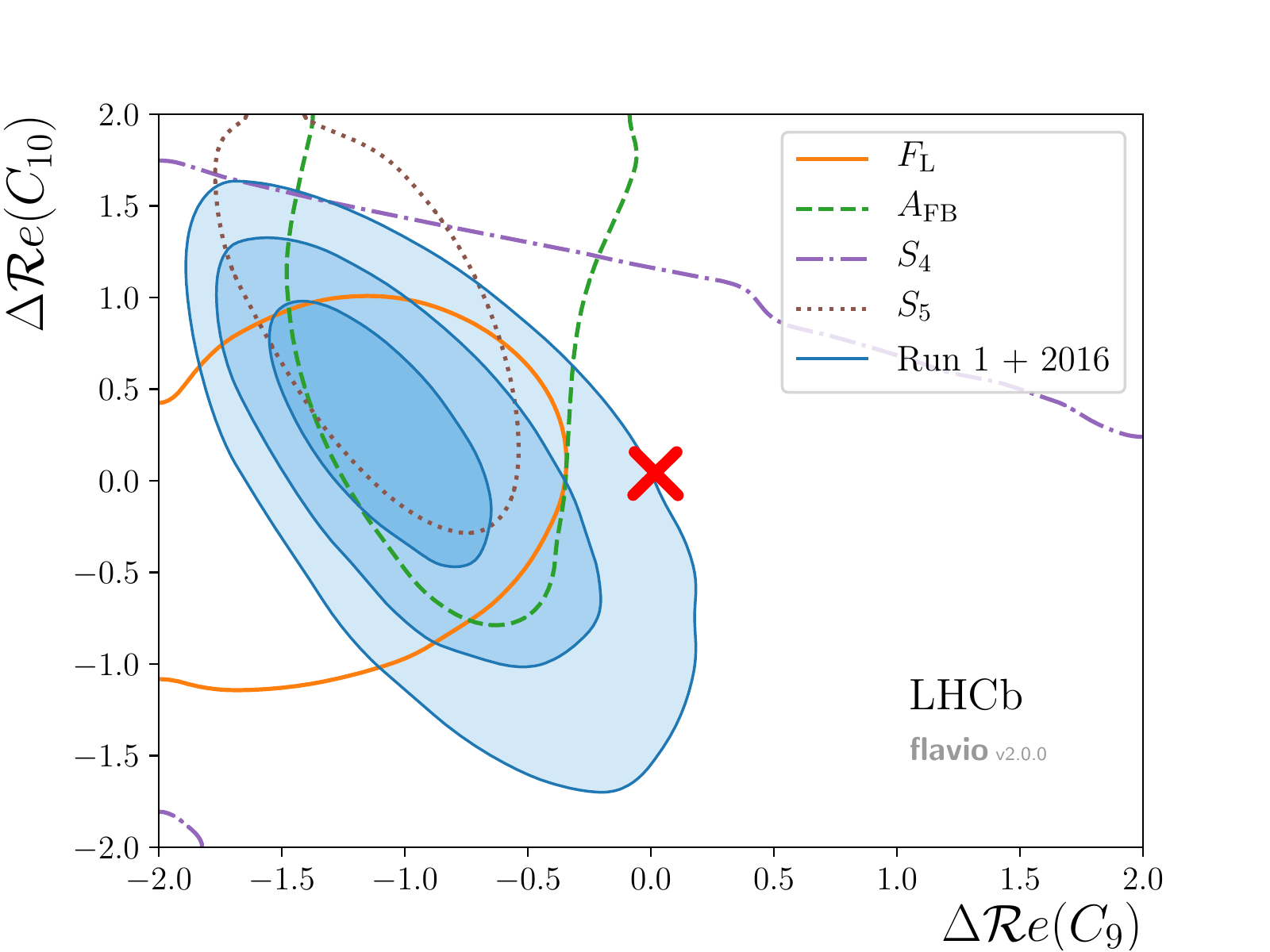}
    \caption{Fits to $\Delta {\rm{Re}}(C_9)$ and $\Delta {\rm{Re}}(C_{10})$, the shift in the values of real part of the Wilson coefficients $C_9$ and $C_{10}$ from their SM values, using the measurements of various angular observables in the decay \decay{\Bdb}{\Kstarzb\mup\mun}. The measurements are consistent with each other and are $\sim 3 \sigma$ from the SM point, which is marked with a red cross. Adapted from Ref.~\cite{LHCb:2020lmf}. }
    \label{fig:C9_C10_Kstarmumu}
\end{figure}

To some greater or lesser extent, angular observables give access to quantities that are independent of the dominant theoretical uncertainties, which come from the hadronic form factors~\cite{Bharucha:2015bzk, Horgan:2013hoa}. While individual angular observables may then enable a clean comparison with SM predictions - for example the quasi form-factor independent observable $P_5^{\prime}$ has been measured to have some local tension with the SM in the region $2.5 < \qsq < 6.0\gevgevcccc$~\cite{LHCb:2020lmf}; the combination of a full basis of angular observables enables the form factors and other theory nuisance parameters to be constrained in global fits. In \decay{\Bdb}{\Kstarzb\mup\mun} decays, angular measurements are then made to the observables $A_{\mathrm{FB}}$, $F_L$ and the $P_i^{\prime}$ series of quantities; or the $S_i$ series of observables~\cite{Kruger:1999xa, Descotes-Genon:2013vna, Altmannshofer:2008dz}; enabling the underlying Wilson coefficients $C_9, C_{10}$ and the theory nuisance parameters to be determined. The majority of measurements in different \qsq bins are in agreement with SM predictions. The local deviation in $P_5^{\prime}$ or $S_5$  observed at LHCb is consistent with smaller effects seen in other observables and can be attributed to a new vector contribution that shifts the $C_9$ Wilson coefficient (see Fig.~\ref{fig:C9_C10_Kstarmumu}). From the angular analysis, there is presently no indication for new axial vector contribution that shifts $C_{10}$.

The dominant theory uncertainty on $P_5^{\prime}$ is from the so-called $c\bar{c}$ loop effect. This includes the interference between the amplitudes for $\decay{\Bdb}{\jpsi(\to \mup\mun)\Kstarzb(\to \Km\pip)}$ and the \decay{\Bdb}{\Kstarzb\mup\mun} process. There is a broad consensus in the theoretical community that the effect from such decays is large in the region $\qsq > 6.0 \gev/c^4$ but the degree to which such effects could influence the measurements made in the region $\qsq < 6.0 \gev/c^4$ is still subject to debate~\cite{Jager:2014rwa,Lyon:2014hpa,Ciuchini:2015qxb, Bobeth:2017vxj}. The relevant theoretical calculations involve non-local effects and hence are intractable. The $c\bar{c}$ processes give a vector-like contribution that could adjust the $C_9$ Wilson coefficient, and hence mimic a new physics contribution. A genuine effect from heavy new physics would be independent of \qsq, whereas the interference from $c\bar{c}$ loops would be dependent on \qsq. Future analyses will try to determine the interference by fitting models for the resonant and rare decay amplitudes to the experimental data unbinned in \qsq~\cite{Blake:2017fyh}.
    
In addition to the \decay{\Bdb}{\Kstarzb\mup\mun} with \decay{\Kstarzb}{\Km\pip}, the \decay{\Bm}{\Kstarm\mup\mun} with \decay{\Kstarm}{\KS\pim}~\cite{LHCb:2020gog} and  \decay{\Bm}{\Km\mup\mun}~\cite{LHCb:2014auh}, angular distributions have been measured by the LHCb collaboration. The former also shows some tension with respect to SM predictions. Although the \decay{\Bm}{\Km\mup\mun} decay, measurements of the $A_{\mathrm FB}$ and $F_H$ angular observables are in good agreement with the SM predictions, there is no inconsistency, as they are only sensitive to new (pseudo)scalar and tensor couplings. Measurements have also been made for the \decay{\Bsb}{\phi\mup\mun}~\cite{LHCb:2021xxq} decay, where the $\phi$ decays to a CP eigenstate and hence does not tag the \B flavour. Measurements of the baryonic decay \decay{\Lb}{\Lambda \mup\mun} have also been performed ~\cite{LHCb:2018jna}. The different spin structure gives access to different physics but the decay  is experimentally much more  challenging given the relatively long lifetime of the $\Lambda$.

Analysis of \decay{\Bdb}{\Kstarzb\ep\en} decays in the  \qsq region dominated by the photon pole (the interval between 0.0008 and 0.257\gevgevcccc) provides the current best constraints on the right-handed currents via the $C_7^{\prime}$ Wilson coefficient by measuring the virtual photon polarization (see Sec.~\ref{sec:TCPV}) with 5\% accuracy~\cite{LHCb:2020dof}.
 
At the \belle experiment, reconstruction of the electron and muons modes is comparably efficient and precise measurements are possible in both cases. The angular analysis of \decay{\Bdb}{\Kstarzb\ep\en} decays has enabled the difference between electron and muon decays to be studied and the observable $Q_5= P_5^{\prime}(\mu) - P_5^{\prime}(e)$~\cite{Capdevila:2016ivx} to be formed, yielding further theoretically pristine tests of the SM, albeit with limited experimental precision~\cite{Belle:2016fev}. The measurements are presently consistent with the SM. Analogous studies are in progress at \lhcb but have lower yields compared to the muon modes due to the trigger and are complicated by the bremsstrahlung from the electrons, which results in larger backgrounds. Nonetheless, measurements of angular observables in \decay{\Bdb}{\Kstarzb\ep\en} and related decays will allow $Q_5$ and further differences between angular observables in muon and electron modes to be investigated in the future.

\section{\texorpdfstring{\CP}{CP} VIOLATION}

\subsection{Direct \texorpdfstring{\CP}{CP} violation}

Direct \CP violation is a phenomenon also sensitive to NP.
Suppose a decay $\B \to f$ can proceed through two diagrams with amplitudes
$A_1$ and $A_2$. The amplitudes can be written as
$A_i = |A_i| \exp(i\theta_i)\exp(i\delta_i)$ where $\theta_i$ and
$\delta_i$ are the weak and strong phases in the diagram $i$, respectively.
Comparing the amplitude $A(\B \to f) = A_1 + A_2$ to the \CP conjugate amplitude $A(\Bb \to \offsetoverline{f}) = \bar{A}_1 + \bar{A}_2$, only the sign of the weak phases
flips while the strong phase stays the same. From this, the direct \CP asymmetry, \ACP, 
is obtained as 
\begin{equation}
 \ACP = \frac{\Gamma(\B \to f) - \Gamma (\Bb \to \offsetoverline{f})}{%
  \Gamma(\B \to f) + \Gamma (\Bb \to \offsetoverline{f})}
  = \frac{2|A_1||A_2|\sin(\theta_1 - \theta_2)\sin(\delta_1 - \delta2)}{%
  |A_1|^2 + |A_2|^2 + 2 |A_1| |A_2|
  \cos(\theta_1 - \theta_2)\cos(\delta_1 - \delta2)},
\end{equation}
where $\Gamma = |A|^2$ is the decay width.
$\ACP$ becomes non zero when $\theta_1 \ne \theta_2$,
$\delta_1 \ne \delta_2$ and $|A_1| \sim |A_2|$.
That is, direct \CP violation occurs when a decay proceeds
through two diagrams with similar amplitudes and
with different weak and strong phases.

For exclusive \btosgam and $\bsll$ decays, $\ACP$ is at the few percent level in the SM.
The predictions are also relatively precise
because the uncertainties from the form factors,
which make the branching fraction calculations difficult,
cancel. For this reason, $\ACP$ of exclusive decays have been intensively
measured as a probe for NP. Presently, $\ACP$ for $\decay{\Bzb}{\Kstarzb\gamma}$ decays is measured to be
$-0.006 \pm 0.011$~\cite{PDG2020}, while the uncertainty
of the theoretical prediction is a few percent.
The asymmetries \ACP for $\bsll$ and $\bquark \to \dquark\ell^+ \ell^-$ decays
have also been measured
with precision of around 10\% for many decays~\cite{PDG2020}
but as good as 2\% for $\Bb \to \Kbar^{(*)}\mup\mun$~\cite{LHCb:2014mit}.
The results are consistent with null asymmetry.

$\ACP$ in the inclusive modes may be more interesting.
The $\ACP$ in $B \to X_s\gamma$ is predicted to be
$-0.006 < \ACP(B \to X_s\gamma) < 0.028$~\cite{Benzke:2010tq}.
Experimentally, this can be measured by reconstructing
$X_s$ as a sum of many exclusive final states, e.g.\ $K\pi$, $K\pi\pi$.
The average of the measurements by the BaBar~\cite{BaBar:2014czi} and Belle~\cite{Belle:2018iff} collaborations is
$\ACP(B \to X_s\gamma) = 0.015 \pm 0.011$~\cite{PDG2020}, which 
is consistent with the SM prediction.
Another variable $\Delta \ACP$, isospin violation of $\ACP$,
defined as $\ACP(B \to X_s^-\gamma) - \ACP(B \to X_s^0\gamma)$
is proposed as a cleaner variable~\cite{Benzke:2010tq} with experimental measurements from the same papers.

Another approach is to measure $\ACP(B \to X \gamma)$ inclusively,
without distinguishing $B \to X_s\gamma$ and $B \to X_d\gamma$.
In the SM, $\ACP$ in \btosgam and \btodgam cancels out,
and $\ACP(B \to X \gamma)$ is expected to be null,
so this provides an excellent test of the SM.
The analysis strategy is the same as the branching fraction measurement.
The flavour (whether the process is
the decay of a \bquark or a \bquarkbar) can be obtained
from the other $B$ in the event, either from a lepton in the semileptonic
decay of the other $B$, or from a reconstructed $B$ meson when the 
other $B$ meson is fully reconstructed.
The probability of the wrong tag of the flavour,
which partially comes from \Bz-\Bzb oscillations, needs to
be taken into account, in the analysis.
The present world average is 
$\ACP(B \to X\gamma) = 0.010 \pm 0.031$~\cite{PDG2020}.

\subsection{Time-dependent \texorpdfstring{\CP}{CP} violation and Polarization in Radiative \texorpdfstring{$B$}{B} decays}
\label{sec:TCPV}

When the $\Bz$ and $\Bzb$ can decay to a common \CP eigenstate $f$,
the process $\Bz \to f$ can interfere with $\Bz \to \Bzb \to f$ process,
causing time-dependent \CP violation to occur with an amplitude $S$.
For example, the time-dependent \CP asymmetry in $\Bzb \to \jpsi\KS$ decays
can be used to determine the angle $\beta$ ($=\phi_1$)
of the unitarity triangle through the $\CP$ violating parameter 
$S = \sin2\beta$.

\begin{marginnote}[]
\entry{Unitarity triangle}{%
In the SM, \CP violation occurs though a complex phase in the CKM matrix elements.
A triangle can be drawn on the complex plane from
the relation of CKM matrix elements, $|\Vud\Vub^* + \Vcd\Vcb^* + \Vtd\Vtb^* =0$, due
to the unitarity of the CKM matrix.
This triangle is called the unitarity triangle, and the three interior angles are named as
$\alpha$ ($=\phi_2$), $\beta$ ($=\phi_1$), $\gamma$ ($=\phi_3)$.
}
\end{marginnote}

Time-dependent \CP violation can be also considered in the 
radiative $B$ decays $\Bzb \to X^0 \gamma$ where $X^0$ is a
neutral hadronic state for which flavour is not identified,
e.g. $\KS\piz$.
In the SM, the photon in \btosgam ($\Bzb \to X_s\gamma$) is
left-handed, while the photon in $\bquarkbar \to \squarkbar\gamma$
($\Bz \to X_{\squarkbar}\gamma$) is right-handed.
Therefore, \Bz and \Bzb do not decay into a common \CP eigenstate
and no time-dependent \CP violation occurs.
To be precise, the right-handed photon is not totally inhibited,
but is suppressed by the quark mass ratio $m_\squark/m_\bquark$,
and $S \sim -2 m_\squark/m_\bquark \sin 2\beta$ is expected
for $\Bzb \to X_s\gamma$.
The same discussion holds for the \btodgam replacing the \squark quark with the \dquark quark,
though in this case the weak phase appearing in $S$ vanishes.
In any case, in radiative $B$ decays, $S$ is proportional to the fraction
of the right-handed current $C'_7$ in \btosgam(\btodgam); hence
the measurement of the time-dependent \CP asymmetry is
essentially the measurement of the photon polarization.

Time-dependent \CP violation in \btosgam has been measured by BaBar and Belle (\eg Refs.~\cite{BaBar:2008okc,Belle:2007kjy}).
For example, $S$ for $B^0 \to \Kstarzb(\to \KS\piz)\gamma$ is measured to be
$-0.15 \pm 0.22$ by the BaBar and Belle collaborations~\cite{PDG2020},
which is consistent with no time-dependent \CP violation.
Measurements have been made with $B^0 \to \KS(\eta,\rho^0,\phi)\gamma$
and $B^0 \to \rho^0\gamma$ with larger uncertainty.
\B factory experiments have the advantage
for these measurement, since 
the final state involves neutral particles, and forming a \KS decay vertex is necessary for many modes.

Similar measurements can be made with $\Bsb \to \phi\gamma$ decays. As a measurement of $S$ requires oscillations to be resolved, a large boost of the \Bsb mesons is required and a measurement is only possible at the \lhcb experiment. Measurements of $S$ and
$\mathcal{A}^\Delta$, which is another 
parameter sensitive to the photon polarization amplitudes, has been made 
with uncertainties of around 30\% using $3~\invfb$ of data~\cite{LHCb:2019vks}.
This is one of the promising modes by \lhcb for the precise study
on the photon polarization.

Other methods to measure the photon polarization in radiative
$B$ decays also exist.
One possibility is to utilize the photon conversion $\gamma \to e^+e^-$
in the detector materials for the decay $B \to K^* (\to K\pi) \gamma$.
However, this method requires very thin materials 
to reduce the effect of the multiple scattering
and it does not look feasible with \lhcb and \belletwo. The $\Bzb \to \Kstarzb e^+e^-$ decay at $0.0008 < \qsq < 0.257\gevgevcccc$ is completely dominated by a virtual photon and the angular distribution can be used to measure the photon polarisation, as discussed in Sec.~\ref{sec:angular}.

\section{LEPTON FLAVOUR UNIVERSALITY AND LEPTON FLAVOUR VIOLATION}
\label{sec:LFU}
Decays of \bquark-hadrons involving a \bTosll transition provide an ideal laboratory to test Lepton Universality (LU) due to the absence of a SM tree-level contribution. They are sensitive to contributions from new particles that can induce a sizeable increase or decrease in the rate of given decays. These contributions can also modify the angular distribution of the final-state particles, which can provide further tests of LU. Whether these tests compare the branching fractions or angular distributions of the two lepton types, they are all characterised by quasi negligible theoretical uncertainties on the SM predictions.

\begin{figure}
  \centering
  \begin{tikzonimage}[height=0.85\textwidth,width=1.\textwidth]{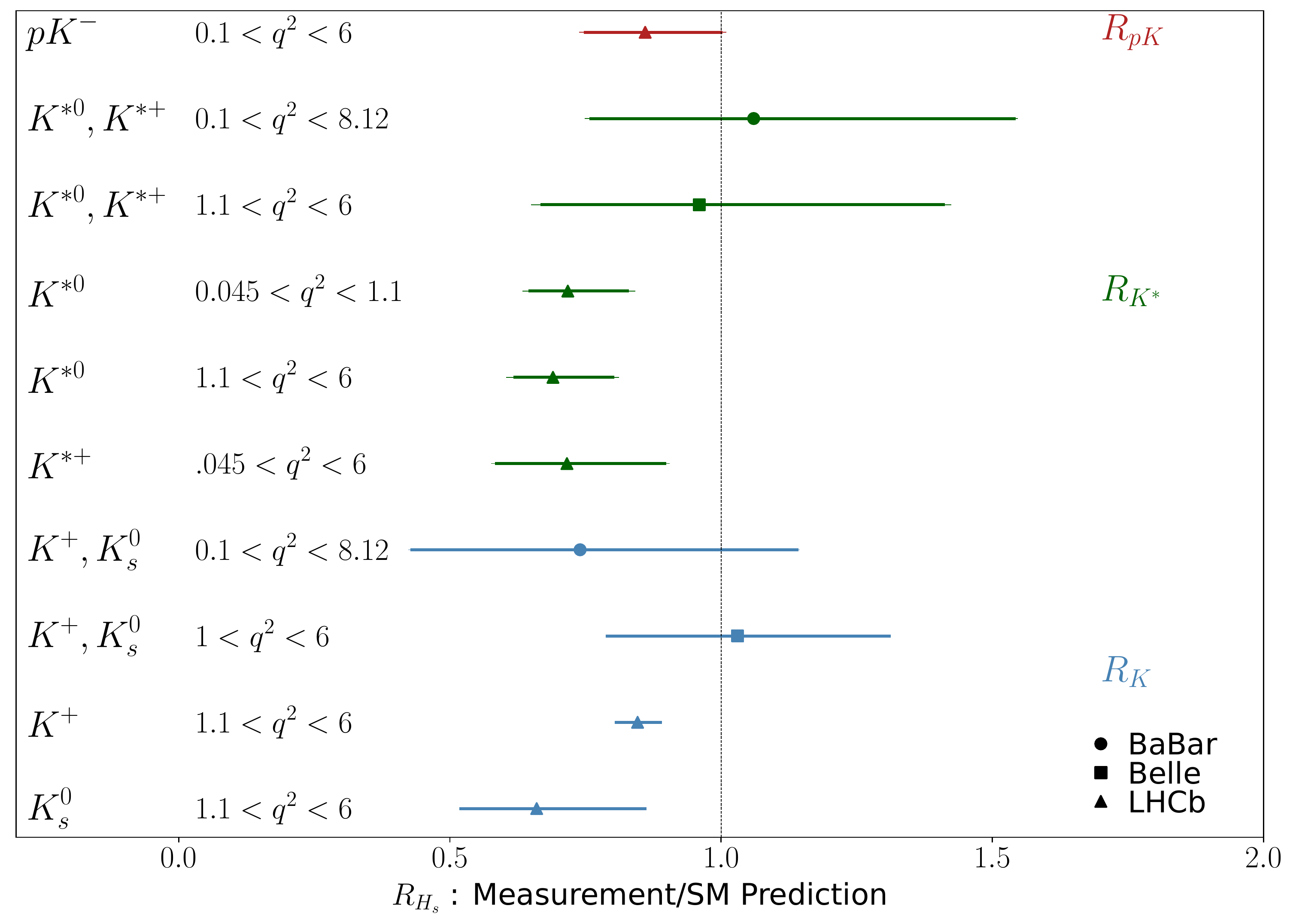}
  \end{tikzonimage}
 \caption{Summary of the results for Lepton Universality tests on ratios of branching fractions. The \qsq intervals are expressed in units of \gevgevcccc.}
 \label{fig:RHs}
\end{figure}

Currently, most results consist in ratios of the type~\cite{Hiller:2003js}
\begin{equation}
R_{H_s} = \frac{\bigintssss_{\qsqmin}^{\qsqmax} \frac{ d\Gamma(H_b \to H_s\mumu) }{d\qsq} \, d\qsq}{\bigintssss_{\qsqmin}^{\qsqmax} \frac{ d\Gamma(H_b \to H_s\epem) }{d\qsq} \, d\qsq} \, ,
\end{equation}
where $H_b$ represents a hadron containing a \bquark-quark, $H_s$ represents a hadron containing an \squark-quark, and \qsq is the invariant mass squared of the dilepton system integrated between \qsqmin and \qsqmax.
In the SM such ratios are expected to be close to unity above the dimuon threshold. 
For all $R_{H_s}$ experimental results the \qsq regions around the charmonium resonances are vetoed since they are dominated by SM physics. 
In the \lhcb case, in order to be less sensitive to the detailed knowledge of the electron and muon reconstruction efficiencies, 
the $H_b \to H_s \lplm$ branching fractions are measured relative to the branching fraction of the $H_b \to H_s \jpsi (\to \lplm) $ decay, exploiting the fact that the $\jpsi \to \lplm$ decay is known to respect LU at the 0.3\% level~\cite{PDG2020}.   
Current results are summarised in Fig.~\ref{fig:RHs}. The \lhcb results are only marginally compatible with unity, and the most precise result~\cite{LHCb-PAPER-2021-004} is 3.1 standard deviations from the SM prediction. The coming years will be very interesting with more results expected from the current and future \lhc datasets, as well as with \belletwo. Indeed \belletwo, with 20\invab expected around 2025, should be able to confirm the \RK measurement from \lhcb. Due to the interest in these measurements, the \atlas\ and \cms\ collaborations are currently investigating their sensitivities. Beyond these improvements in sensitivity and the interest in having very different experimental conditions and thus different experimental systematic uncertainties, the next major steps will come from LU tests using angular analyses as pioneered by \belle~\cite{Wehle:2016yoi}. 

Linked with the issue of possible LU violation another interesting aspect is the search for Charged Lepton Flavour Violation (CLFV) decays. 
In the SM, these decays are completely negligible and an observation would be an indisputable sign of New Physics.  
For \decay{H_b}{\ell\ell'} decays, for ($\Pe , \Pmu$) pairs, the results are dominated by the LHCb results with upper limits of the order of $10^{-9}$ at 90\% CL~\cite{PDG2020}. For the other combinations, more challenging from an experimental point of view, the cleanliness of the \B-factories environment compensates for their lower $\bquark$-hadron production rate. The upper limits are nevertheless only of order $10^{-5}$ at 90 \% CL.
On a similar topic, many searches have been carried out using \decay{H_b}{H_s \ell\ell'} decays. The upper limits follow the same experimental schema as the \decay{H_b}{\ell\ell'} decays: the best limits (of order $10^{-9}$) are obtained by the \lhcb\ experiment for final states with an electron and a muon, while for the other cases, the limits span from $10^{-5}$ to a few $10^{-8}$, with significant contributions from the \babar\ and Belle experiments~\cite{PDG2020}. All these results constrain the broad set of New Physics models proposed to explain the tensions observed in analyses dealing with $\btosll$ transitions, including in some cases the results of LU tests in $b \to c \ell \nu$ decays, in particular those involving a tau lepton~\cite{PDG2020}. 

\section{DISCUSSION}
Since the first measurement of an electroweak penguin decay by the \cleo collaboration in 1993, there are now around 200 different measurements. With the parallel development of the theoretical understanding, the area has moved from something that verified the existence of these type of decays in the SM to an area that is pushing the boundaries for the discovery of New Physics. 

\begin{figure}
    \includegraphics[width=\linewidth]{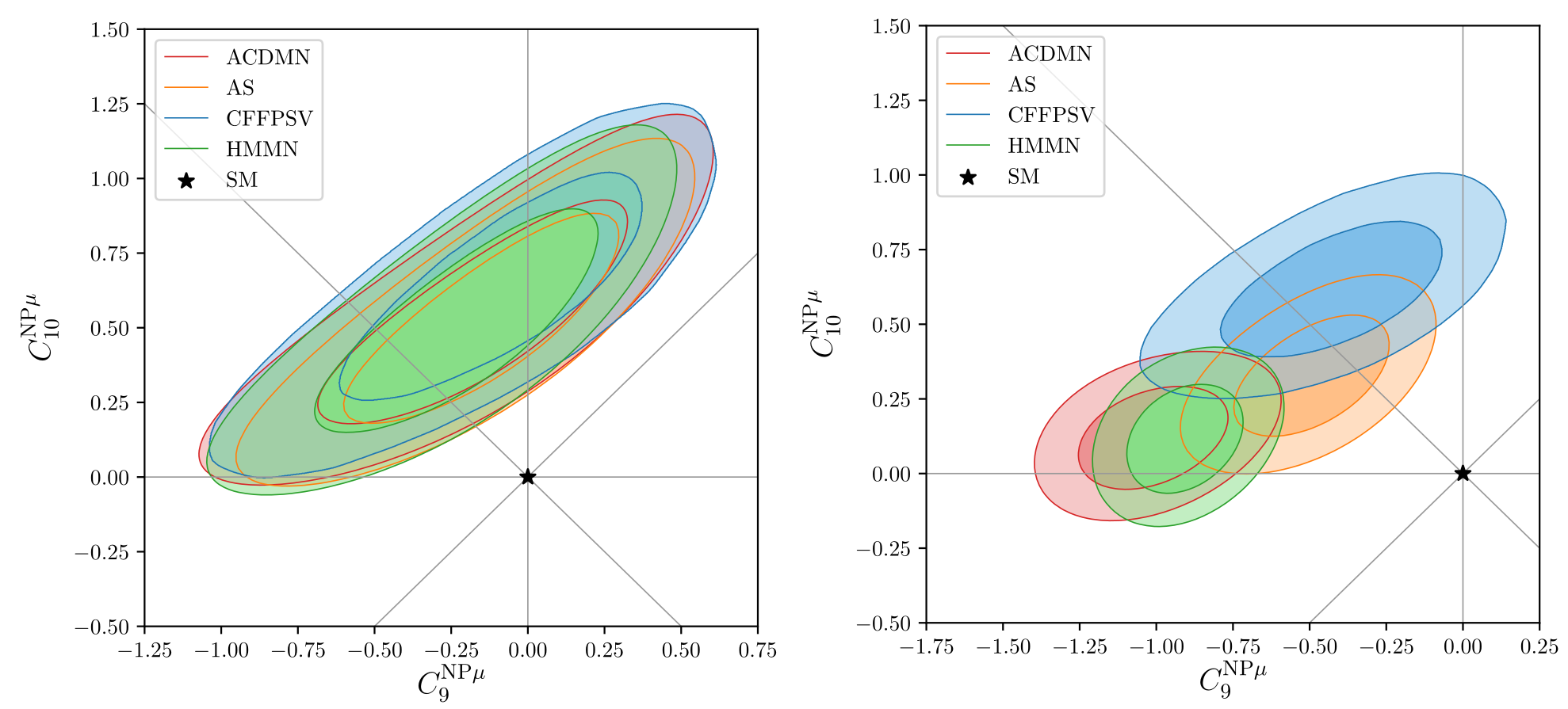}
    \caption{Global fits (ACDMN~\cite{Alguero:2021anc}, AS~\cite{Altmannshofer:2021qrr}, CFFPSV~\cite{Ciuchini:2020gvn}, HMMN~\cite{Hurth:2021nsi}) for any New Physics contributions to the Wilson coefficients $C_9$ and $C_{10}$ under the assumption that all other Wilson coefficients take on their SM values. The fits use (left) lepton flavour universality observations and \decay{\Bsb}{\mup\mun} and (right) all observables. The former results in a poorer experimental resolution but has fewer theoretical assumptions compared to the latter.
    The two contours for each fit correspond to $1\sigma$ and $2\sigma$ uncertainties.}
    \label{fig:globalfit}
\end{figure}

In Fig.~\ref{fig:globalfit} is a comparison of several global fits for Wilson coefficients~\cite{Alguero:2021anc, Altmannshofer:2021qrr, Ciuchini:2020gvn, Hurth:2021nsi}. Current results favour a substantial NP contribution to the Wilson coefficients $C_9$ and/or $C_{10}$ for muons in the final state. The fits represented here assume that all other Wilson coefficients take on their SM values. In the first one, only the observables with high theoretical precision (the lepton universality measurements and  the \decay{\Bs}{\mup\mun} branching fraction) are used, while in the second figure all observables are used. It can be seen that adding more experimental observations results in smaller uncertainties but also a higher level of model dependence, \ie different global fits obtain somewhat different results even if they use the same experimental data.
\begin{marginnote}[]
\entry{Global fit}{A fit that combines many different measurements of $\btosll$ observables to find the most likely values of the Wilson coefficients that would result in those measurements. They involve the combination of experimental measurements, theoretical uncertainties and their respective correlations.}
\end{marginnote}

The \lhcb upgrades and \belletwo will make more accurate measurements in the future and the collaborations have made specific predictions of the accuracy~\cite{LHCb-PII-Physics,Belle-II:2018jsg}.
In addition there are prospects of making entirely new measurements that can either pin down theoretical uncertainties or further investigate the nature of any such New Physics.

The experimental challenge for the future will be to make measurements of electroweak penguin decays in as many different ways as possible as this will decrease the overall statistical uncertainty for global fits but more importantly will explore theoretical assumptions and further clarify the nature of any New Physics. As highlighted in this review, keeping a broad physics programme for the analyses across multiple experiments is of utmost importance to exploit the different experimental constraints, as well as serving as an important cross check for the systematic uncertainties that to some extent are shared between different measurements at the same experiment.

\begin{summary}[SUMMARY POINTS]
\begin{enumerate}
\item Electroweak penguin decays are highly sensitive to New Physics at mass scales well above the electroweak energy scale.
\item Measurements at the \B factories and at the \lhc have made several hundred measurements that through global fits give an indication for New Physics in vector and/or axial vector couplings.
\item Particle identification of leptons, hadrons and photons is essential for performing the measurements.
\item To reduce theoretical uncertainties in the interpretation of results there is a need to perform new measurements where data can be used to minimise these uncertainties.
\end{enumerate}
\end{summary}

\begin{issues}[FUTURE ISSUES]
\begin{enumerate}
    \item Measurements with $\tau$ leptons in the final state would add much to the exploration of lepton flavour universality and lepton number conservation. While measurements of some of these final states have already been made, they are in general still far away from setting serious constraints on NP. An exciting idea is the measurement of new observables in the decay \decay{\Bdb}{\Kstarzb\taup\taum} in the far future at FCC-ee~\cite{Kamenik:2017ghi}.
    \item At the end of the high luminosity LHC period, it will be possible to make a measurement of the ratio of branching fractions for the decays \decay{\Bdb}{\mup\mun} and \decay{\Bsb}{\mup\mun}~\cite{ATL-PHYS-PUB-2018-005, CMS-PAS-FTR-18-013, LHCb-PII-Physics}. Such a measurement explores the minimal flavour violation paradigm of New Physics with almost no theoretical uncertainty.
    \item High precision simultaneous angular analyses of the decays \decay{\Bdb}{\Kstarzb\mup\mun} and \decay{\Bdb}{\Kstarzb\epem} will allow for a precision determination of Lepton Flavour Violating NP. This comes from the fact that the hadronic uncertainties are shared between the modes and thus will cancel out in a comparison. 
\end{enumerate}
\end{issues}

\section*{DISCLOSURE STATEMENT}
UE, MP and MHS are members of the \lhcb collaboration. SN is a member of the \belle and \belletwo collaborations.

\section*{ACKNOWLEDGMENTS}
UE would like to acknowledge funding from ARC grant DP210102707.
SN would like to acknowledge funding from JSPS KAKENHI Grant Number JP17K05474. 
MHS would like to acknowledge funding from European Research Council (ERC) under the European Union’s Horizon 2020 research and innovation programme (grant agreement No 101018181). MP acknowledges support from the UK Science and Technology Facilities Council.


\addcontentsline{toc}{section}{References}
\bibliographystyle{LHCb}
\bibliography{main,standard,LHCb-PAPER,LHCb-DP}

\ifx\mcitethebibliography\mciteundefinedmacro
\PackageError{LHCb.bst}{mciteplus.sty has not been loaded}
{This bibstyle requires the use of the mciteplus package.}\fi
\providecommand{\href}[2]{#2}
\begin{mcitethebibliography}{10}
\mciteSetBstSublistMode{n}
\mciteSetBstMaxWidthForm{subitem}{\alph{mcitesubitemcount})}
\mciteSetBstSublistLabelBeginEnd{\mcitemaxwidthsubitemform\space}
{\relax}{\relax}

\bibitem{Glashow:1970gm}
S.~L. Glashow, J.~Iliopoulos, and L.~Maiani,
  \ifthenelse{\boolean{articletitles}}{\emph{{Weak Interactions with
  Lepton-Hadron Symmetry}},
  }{}\href{https://doi.org/10.1103/PhysRevD.2.1285}{Phys.\ Rev.\  \textbf{D2}
  (1970) 1285}\relax
\mciteBstWouldAddEndPuncttrue
\mciteSetBstMidEndSepPunct{\mcitedefaultmidpunct}
{\mcitedefaultendpunct}{\mcitedefaultseppunct}\relax
\EndOfBibitem
\bibitem{Shifman:1995hc}
M.~A. Shifman, \ifthenelse{\boolean{articletitles}}{\emph{{Foreword to ITEP
  lectures in particle physics}},
  }{}\href{http://arxiv.org/abs/hep-ph/9510397}{{\normalfont\ttfamily
  arXiv:hep-ph/9510397}}\relax
\mciteBstWouldAddEndPuncttrue
\mciteSetBstMidEndSepPunct{\mcitedefaultmidpunct}
{\mcitedefaultendpunct}{\mcitedefaultseppunct}\relax
\EndOfBibitem
\bibitem{Mannel:2004ce}
T.~Mannel, \ifthenelse{\boolean{articletitles}}{\emph{{Effective Field Theories
  in Flavor Physics}}, }{}\href{https://doi.org/10.1007/b62268}{Springer Tracts
  Mod.\ Phys.\  \textbf{203} (2004) }\relax
\mciteBstWouldAddEndPuncttrue
\mciteSetBstMidEndSepPunct{\mcitedefaultmidpunct}
{\mcitedefaultendpunct}{\mcitedefaultseppunct}\relax
\EndOfBibitem
\bibitem{DAmbrosio:2002vsn}
G.~D'Ambrosio, G.~F. Giudice, G.~Isidori, and A.~Strumia,
  \ifthenelse{\boolean{articletitles}}{\emph{{Minimal flavor violation: An
  Effective field theory approach}},
  }{}\href{https://doi.org/10.1016/S0550-3213(02)00836-2}{Nucl.\ Phys.\
  \textbf{B645} (2002) 155},
  \href{http://arxiv.org/abs/hep-ph/0207036}{{\normalfont\ttfamily
  arXiv:hep-ph/0207036}}\relax
\mciteBstWouldAddEndPuncttrue
\mciteSetBstMidEndSepPunct{\mcitedefaultmidpunct}
{\mcitedefaultendpunct}{\mcitedefaultseppunct}\relax
\EndOfBibitem
\bibitem{HFLAV18}
Heavy Flavor Averaging Group, Y.~Amhis {\em et~al.},
  \ifthenelse{\boolean{articletitles}}{\emph{{Averages of $b$-hadron,
  $c$-hadron, and $\tau$-lepton properties as of 2018}},
  }{}\href{https://doi.org/10.1140/epjc/s10052-020-8156-7}{Eur.\ Phys.\ J.\
  \textbf{C81} (2021) 226},
  \href{http://arxiv.org/abs/1909.12524}{{\normalfont\ttfamily
  arXiv:1909.12524}}, {updated results and plots available at
  \href{https://hflav.web.cern.ch}{{\texttt{https://hflav.web.cern.ch}}}}\relax
\mciteBstWouldAddEndPuncttrue
\mciteSetBstMidEndSepPunct{\mcitedefaultmidpunct}
{\mcitedefaultendpunct}{\mcitedefaultseppunct}\relax
\EndOfBibitem
\bibitem{PDG2020}
Particle Data Group, P.~A. Zyla {\em et~al.},
  \ifthenelse{\boolean{articletitles}}{\emph{{\href{http://pdg.lbl.gov/}{Review
  of particle physics}}}, }{}\href{https://doi.org/10.1093/ptep/ptaa104}{Prog.\
  Theor.\ Exp.\ Phys.\  \textbf{2020} (2020) 083C01}\relax
\mciteBstWouldAddEndPuncttrue
\mciteSetBstMidEndSepPunct{\mcitedefaultmidpunct}
{\mcitedefaultendpunct}{\mcitedefaultseppunct}\relax
\EndOfBibitem
\bibitem{LHCb-PII-Physics}
LHCb collaboration, \ifthenelse{\boolean{articletitles}}{\emph{{Physics case
  for an LHCb Upgrade II --- Opportunities in flavour physics, and beyond, in
  the HL-LHC era}},
  }{}\href{http://arxiv.org/abs/1808.08865}{{\normalfont\ttfamily
  arXiv:1808.08865}}\relax
\mciteBstWouldAddEndPuncttrue
\mciteSetBstMidEndSepPunct{\mcitedefaultmidpunct}
{\mcitedefaultendpunct}{\mcitedefaultseppunct}\relax
\EndOfBibitem
\bibitem{LHCb-PAPER-2017-013}
LHCb collaboration, R.~Aaij {\em et~al.},
  \ifthenelse{\boolean{articletitles}}{\emph{{Test of lepton universality with
  \mbox{\decay{\Bz}{\Kstarz\ellell}} decays}},
  }{}\href{https://doi.org/10.1007/JHEP08(2017)055}{JHEP \textbf{08} (2017)
  055}, \href{http://arxiv.org/abs/1705.05802}{{\normalfont\ttfamily
  arXiv:1705.05802}}\relax
\mciteBstWouldAddEndPuncttrue
\mciteSetBstMidEndSepPunct{\mcitedefaultmidpunct}
{\mcitedefaultendpunct}{\mcitedefaultseppunct}\relax
\EndOfBibitem
\bibitem{Abe:2002rc}
Belle collaboration, K.~Abe {\em et~al.},
  \ifthenelse{\boolean{articletitles}}{\emph{{Measurement of branching
  fractions and charge asymmetries for two-body B meson decays with
  charmonium}}, }{}\href{https://doi.org/10.1103/PhysRevD.67.032003}{Phys.\
  Rev.\  \textbf{D67} (2003) 032003},
  \href{http://arxiv.org/abs/hep-ex/0211047}{{\normalfont\ttfamily
  arXiv:hep-ex/0211047}}\relax
\mciteBstWouldAddEndPuncttrue
\mciteSetBstMidEndSepPunct{\mcitedefaultmidpunct}
{\mcitedefaultendpunct}{\mcitedefaultseppunct}\relax
\EndOfBibitem
\bibitem{LHCb:2018zdd}
LHCb collaboration, R.~Aaij {\em et~al.},
  \ifthenelse{\boolean{articletitles}}{\emph{{Design and performance of the
  LHCb trigger and full real-time reconstruction in Run 2 of the LHC}},
  }{}\href{https://doi.org/10.1088/1748-0221/14/04/P04013}{JINST \textbf{14}
  (2019) P04013}, \href{http://arxiv.org/abs/1812.10790}{{\normalfont\ttfamily
  arXiv:1812.10790}}\relax
\mciteBstWouldAddEndPuncttrue
\mciteSetBstMidEndSepPunct{\mcitedefaultmidpunct}
{\mcitedefaultendpunct}{\mcitedefaultseppunct}\relax
\EndOfBibitem
\bibitem{LHCb:2021qbv}
LHCb collaboration, R.~Aaij {\em et~al.},
  \ifthenelse{\boolean{articletitles}}{\emph{{Precise measurement of
  the~$f_s/f_d$ ratio of fragmentation fractions and of $B^0_s$ decay branching
  fractions}}, }{}\href{https://doi.org/10.1103/PhysRevD.104.032005}{Phys.\
  Rev.\  \textbf{D104} (2021) 032005},
  \href{http://arxiv.org/abs/2103.06810}{{\normalfont\ttfamily
  arXiv:2103.06810}}\relax
\mciteBstWouldAddEndPuncttrue
\mciteSetBstMidEndSepPunct{\mcitedefaultmidpunct}
{\mcitedefaultendpunct}{\mcitedefaultseppunct}\relax
\EndOfBibitem
\bibitem{Bobeth:2013uxa}
C.~Bobeth {\em et~al.}, \ifthenelse{\boolean{articletitles}}{\emph{{$B_{s,d}
  \to l^+ l^-$ in the Standard Model with Reduced Theoretical Uncertainty}},
  }{}\href{https://doi.org/10.1103/PhysRevLett.112.101801}{Phys.\ Rev.\ Lett.\
  \textbf{112} (2014) 101801},
  \href{http://arxiv.org/abs/1311.0903}{{\normalfont\ttfamily
  arXiv:1311.0903}}\relax
\mciteBstWouldAddEndPuncttrue
\mciteSetBstMidEndSepPunct{\mcitedefaultmidpunct}
{\mcitedefaultendpunct}{\mcitedefaultseppunct}\relax
\EndOfBibitem
\bibitem{Beneke:2019slt}
M.~Beneke, C.~Bobeth, and R.~Szafron,
  \ifthenelse{\boolean{articletitles}}{\emph{{Power-enhanced
  leading-logarithmic QED corrections to $B_q \to \mu^+\mu^-$}},
  }{}\href{https://doi.org/10.1007/JHEP10(2019)232}{JHEP \textbf{10} (2019)
  232}, \href{http://arxiv.org/abs/1908.07011}{{\normalfont\ttfamily
  arXiv:1908.07011}}\relax
\mciteBstWouldAddEndPuncttrue
\mciteSetBstMidEndSepPunct{\mcitedefaultmidpunct}
{\mcitedefaultendpunct}{\mcitedefaultseppunct}\relax
\EndOfBibitem
\bibitem{ATLAS:2018cur}
ATLAS collaboration, M.~Aaboud {\em et~al.},
  \ifthenelse{\boolean{articletitles}}{\emph{{Study of the rare decays of
  $B^0_s$ and $B^0$ mesons into muon pairs using data collected during 2015 and
  2016 with the ATLAS detector}},
  }{}\href{https://doi.org/10.1007/JHEP04(2019)098}{JHEP \textbf{04} (2019)
  098}, \href{http://arxiv.org/abs/1812.03017}{{\normalfont\ttfamily
  arXiv:1812.03017}}\relax
\mciteBstWouldAddEndPuncttrue
\mciteSetBstMidEndSepPunct{\mcitedefaultmidpunct}
{\mcitedefaultendpunct}{\mcitedefaultseppunct}\relax
\EndOfBibitem
\bibitem{CMS:2019bbr}
CMS collaboration, A.~M. Sirunyan {\em et~al.},
  \ifthenelse{\boolean{articletitles}}{\emph{{Measurement of properties of
  B$^0_\mathrm{s}\to\mu^+\mu^-$ decays and search for B$^0\to\mu^+\mu^-$ with
  the CMS experiment}}, }{}\href{https://doi.org/10.1007/JHEP04(2020)188}{JHEP
  \textbf{04} (2020) 188},
  \href{http://arxiv.org/abs/1910.12127}{{\normalfont\ttfamily
  arXiv:1910.12127}}\relax
\mciteBstWouldAddEndPuncttrue
\mciteSetBstMidEndSepPunct{\mcitedefaultmidpunct}
{\mcitedefaultendpunct}{\mcitedefaultseppunct}\relax
\EndOfBibitem
\bibitem{LHCb:2021vsc}
LHCb collaboration, R.~Aaij {\em et~al.},
  \ifthenelse{\boolean{articletitles}}{\emph{{Analysis of neutral $B$-meson
  decays into two muons}},
  }{}\href{http://arxiv.org/abs/2108.09284}{{\normalfont\ttfamily
  arXiv:2108.09284}}\relax
\mciteBstWouldAddEndPuncttrue
\mciteSetBstMidEndSepPunct{\mcitedefaultmidpunct}
{\mcitedefaultendpunct}{\mcitedefaultseppunct}\relax
\EndOfBibitem
\bibitem{LHCb-PAPER-2014-049}
CMS and LHCb collaborations, V.~Khachatryan {\em et~al.},
  \ifthenelse{\boolean{articletitles}}{\emph{{Observation of the rare
  \mbox{\decay{\Bs}{\mumu}} decay from the combined analysis of CMS and LHCb
  data}}, }{}\href{https://doi.org/10.1038/nature14474}{Nature \textbf{522}
  (2015) 68}, \href{http://arxiv.org/abs/1411.4413}{{\normalfont\ttfamily
  arXiv:1411.4413}}\relax
\mciteBstWouldAddEndPuncttrue
\mciteSetBstMidEndSepPunct{\mcitedefaultmidpunct}
{\mcitedefaultendpunct}{\mcitedefaultseppunct}\relax
\EndOfBibitem
\bibitem{LHCb:2021awg}
LHCb collaboration, R.~Aaij {\em et~al.},
  \ifthenelse{\boolean{articletitles}}{\emph{{Measurement of the
  $B^0_s\to\mu^+\mu^-$ decay properties and search for the $B^0\to\mu^+\mu^-$
  and $B^0_s\to\mu^+\mu^-\gamma$ decays}},
  }{}\href{http://arxiv.org/abs/2108.09283}{{\normalfont\ttfamily
  arXiv:2108.09283}}\relax
\mciteBstWouldAddEndPuncttrue
\mciteSetBstMidEndSepPunct{\mcitedefaultmidpunct}
{\mcitedefaultendpunct}{\mcitedefaultseppunct}\relax
\EndOfBibitem
\bibitem{Ammar:1993sh}
CLEO Collaboration, R.~Ammar {\em et~al.},
  \ifthenelse{\boolean{articletitles}}{\emph{Evidence for penguins: First
  observation of {$B \to K^*(892) \gamma$}},
  }{}\href{https://doi.org/10.1103/PhysRevLett.71.674}{Phys.\ Rev.\ Lett.\
  \textbf{71} (1993) 674}\relax
\mciteBstWouldAddEndPuncttrue
\mciteSetBstMidEndSepPunct{\mcitedefaultmidpunct}
{\mcitedefaultendpunct}{\mcitedefaultseppunct}\relax
\EndOfBibitem
\bibitem{Aaij:2019hhx}
LHCb collaboration, R.~Aaij {\em et~al.},
  \ifthenelse{\boolean{articletitles}}{\emph{{First observation of the
  radiative decay $\Lambda_{b}^{0} \to \Lambda \gamma$}},
  }{}\href{https://doi.org/10.1103/PhysRevLett.123.031801}{Phys.\ Rev.\ Lett.\
  \textbf{123} (2019) 031801},
  \href{http://arxiv.org/abs/1904.06697}{{\normalfont\ttfamily
  arXiv:1904.06697}}\relax
\mciteBstWouldAddEndPuncttrue
\mciteSetBstMidEndSepPunct{\mcitedefaultmidpunct}
{\mcitedefaultendpunct}{\mcitedefaultseppunct}\relax
\EndOfBibitem
\bibitem{Misiak:2015xwa}
M.~Misiak {\em et~al.}, \ifthenelse{\boolean{articletitles}}{\emph{{Updated
  NNLO QCD predictions for the weak radiative B-meson decays}},
  }{}\href{https://doi.org/10.1103/PhysRevLett.114.221801}{Phys.\ Rev.\ Lett.\
  \textbf{114} (2015) 221801},
  \href{http://arxiv.org/abs/1503.01789}{{\normalfont\ttfamily
  arXiv:1503.01789}}\relax
\mciteBstWouldAddEndPuncttrue
\mciteSetBstMidEndSepPunct{\mcitedefaultmidpunct}
{\mcitedefaultendpunct}{\mcitedefaultseppunct}\relax
\EndOfBibitem
\bibitem{Limosani:2009qg}
Belle collaboration, A.~Limosani {\em et~al.},
  \ifthenelse{\boolean{articletitles}}{\emph{{Measurement of Inclusive
  Radiative B-meson Decays with a Photon Energy Threshold of 1.7-GeV}},
  }{}\href{https://doi.org/10.1103/PhysRevLett.103.241801}{Phys.\ Rev.\ Lett.\
  \textbf{103} (2009) 241801},
  \href{http://arxiv.org/abs/0907.1384}{{\normalfont\ttfamily
  arXiv:0907.1384}}\relax
\mciteBstWouldAddEndPuncttrue
\mciteSetBstMidEndSepPunct{\mcitedefaultmidpunct}
{\mcitedefaultendpunct}{\mcitedefaultseppunct}\relax
\EndOfBibitem
\bibitem{Bosch:2004th}
S.~W. Bosch, B.~O. Lange, M.~Neubert, and G.~Paz,
  \ifthenelse{\boolean{articletitles}}{\emph{{Factorization and shape function
  effects in inclusive B meson decays}},
  }{}\href{https://doi.org/10.1016/j.nuclphysb.2004.07.041}{Nucl.\ Phys.\
  \textbf{B699} (2004) 335},
  \href{http://arxiv.org/abs/hep-ph/0402094}{{\normalfont\ttfamily
  arXiv:hep-ph/0402094}}\relax
\mciteBstWouldAddEndPuncttrue
\mciteSetBstMidEndSepPunct{\mcitedefaultmidpunct}
{\mcitedefaultendpunct}{\mcitedefaultseppunct}\relax
\EndOfBibitem
\bibitem{BaBar:2007yhb}
BaBar collaboration, B.~Aubert {\em et~al.},
  \ifthenelse{\boolean{articletitles}}{\emph{{Measurement of the $B \to X_s
  \gamma$ branching fraction and photon energy spectrum using the recoil
  method}}, }{}\href{https://doi.org/10.1103/PhysRevD.77.051103}{Phys.\ Rev.\
  \textbf{D77} (2008) 051103},
  \href{http://arxiv.org/abs/0711.4889}{{\normalfont\ttfamily
  arXiv:0711.4889}}\relax
\mciteBstWouldAddEndPuncttrue
\mciteSetBstMidEndSepPunct{\mcitedefaultmidpunct}
{\mcitedefaultendpunct}{\mcitedefaultseppunct}\relax
\EndOfBibitem
\bibitem{Belle:2005grh}
Belle collaboration, K.~Abe {\em et~al.},
  \ifthenelse{\boolean{articletitles}}{\emph{{Observation of $b \to d \gamma$
  and determination of $|\Vtd/\Vts|$}},
  }{}\href{https://doi.org/10.1103/PhysRevLett.96.221601}{Phys.\ Rev.\ Lett.\
  \textbf{96} (2006) 221601},
  \href{http://arxiv.org/abs/hep-ex/0506079}{{\normalfont\ttfamily
  arXiv:hep-ex/0506079}}\relax
\mciteBstWouldAddEndPuncttrue
\mciteSetBstMidEndSepPunct{\mcitedefaultmidpunct}
{\mcitedefaultendpunct}{\mcitedefaultseppunct}\relax
\EndOfBibitem
\bibitem{CDF:2006imy}
CDF collaboration, A.~Abulencia {\em et~al.},
  \ifthenelse{\boolean{articletitles}}{\emph{{Observation of $\Bs - \Bsb$
  oscillations}},
  }{}\href{https://doi.org/10.1103/PhysRevLett.97.242003}{Phys.\ Rev.\ Lett.\
  \textbf{97} (2006) 242003},
  \href{http://arxiv.org/abs/hep-ex/0609040}{{\normalfont\ttfamily
  arXiv:hep-ex/0609040}}\relax
\mciteBstWouldAddEndPuncttrue
\mciteSetBstMidEndSepPunct{\mcitedefaultmidpunct}
{\mcitedefaultendpunct}{\mcitedefaultseppunct}\relax
\EndOfBibitem
\bibitem{Aaij:2021pkz}
LHCb collaboration, R.~Aaij {\em et~al.},
  \ifthenelse{\boolean{articletitles}}{\emph{{Branching Fraction Measurements
  of the Rare $B^0_s\rightarrow\phi\mu^+\mu^-$ and $B^0_s\rightarrow
  f_2^\prime(1525)\mu^+\mu^-$- Decays}},
  }{}\href{https://doi.org/10.1103/PhysRevLett.127.151801}{Phys.\ Rev.\ Lett.\
  \textbf{127} (2021) 151801},
  \href{http://arxiv.org/abs/2105.14007}{{\normalfont\ttfamily
  arXiv:2105.14007}}\relax
\mciteBstWouldAddEndPuncttrue
\mciteSetBstMidEndSepPunct{\mcitedefaultmidpunct}
{\mcitedefaultendpunct}{\mcitedefaultseppunct}\relax
\EndOfBibitem
\bibitem{LHCb:2016due}
LHCb collaboration, R.~Aaij {\em et~al.},
  \ifthenelse{\boolean{articletitles}}{\emph{{Measurement of the phase
  difference between short- and long-distance amplitudes in the $B^{+}\to
  K^{+}\mu^{+}\mu^{-}$ decay}},
  }{}\href{https://doi.org/10.1140/epjc/s10052-017-4703-2}{Eur.\ Phys.\ J.\
  \textbf{C77} (2017) 161},
  \href{http://arxiv.org/abs/1612.06764}{{\normalfont\ttfamily
  arXiv:1612.06764}}\relax
\mciteBstWouldAddEndPuncttrue
\mciteSetBstMidEndSepPunct{\mcitedefaultmidpunct}
{\mcitedefaultendpunct}{\mcitedefaultseppunct}\relax
\EndOfBibitem
\bibitem{LHCb:2014cxe}
LHCb collaboration, R.~Aaij {\em et~al.},
  \ifthenelse{\boolean{articletitles}}{\emph{{Differential branching fractions
  and isospin asymmetries of $B \to K^{(*)} \mu^+ \mu^-$ decays}},
  }{}\href{https://doi.org/10.1007/JHEP06(2014)133}{JHEP \textbf{06} (2014)
  133}, \href{http://arxiv.org/abs/1403.8044}{{\normalfont\ttfamily
  arXiv:1403.8044}}\relax
\mciteBstWouldAddEndPuncttrue
\mciteSetBstMidEndSepPunct{\mcitedefaultmidpunct}
{\mcitedefaultendpunct}{\mcitedefaultseppunct}\relax
\EndOfBibitem
\bibitem{BaBar:2013npw}
BaBar collaboration, J.~P. Lees {\em et~al.},
  \ifthenelse{\boolean{articletitles}}{\emph{{Search for $B \to K^{(*)} \nu
  \overline \nu$ and invisible quarkonium decays}},
  }{}\href{https://doi.org/10.1103/PhysRevD.87.112005}{Phys.\ Rev.\
  \textbf{D87} (2013) 112005},
  \href{http://arxiv.org/abs/1303.7465}{{\normalfont\ttfamily
  arXiv:1303.7465}}\relax
\mciteBstWouldAddEndPuncttrue
\mciteSetBstMidEndSepPunct{\mcitedefaultmidpunct}
{\mcitedefaultendpunct}{\mcitedefaultseppunct}\relax
\EndOfBibitem
\bibitem{Belle-II:2021rof}
Belle-II collaboration, F.~Abudin\'en {\em et~al.},
  \ifthenelse{\boolean{articletitles}}{\emph{{Search for $B^+ \to
  K^+\nu\bar\nu$ decays using an inclusive tagging method at Belle II}},
  }{}\href{https://doi.org/10.1103/PhysRevLett.127.181802}{Phys.\ Rev.\ Lett.\
  \textbf{127} (2021) 181802},
  \href{http://arxiv.org/abs/2104.12624}{{\normalfont\ttfamily
  arXiv:2104.12624}}\relax
\mciteBstWouldAddEndPuncttrue
\mciteSetBstMidEndSepPunct{\mcitedefaultmidpunct}
{\mcitedefaultendpunct}{\mcitedefaultseppunct}\relax
\EndOfBibitem
\bibitem{Bause:2021ply}
R.~Bause, H.~Gisbert, M.~Golz, and G.~Hiller,
  \ifthenelse{\boolean{articletitles}}{\emph{{Interplay of dineutrino modes
  with semileptonic rare ${B}$-decays}},
  }{}\href{http://arxiv.org/abs/2109.01675}{{\normalfont\ttfamily
  arXiv:2109.01675}}\relax
\mciteBstWouldAddEndPuncttrue
\mciteSetBstMidEndSepPunct{\mcitedefaultmidpunct}
{\mcitedefaultendpunct}{\mcitedefaultseppunct}\relax
\EndOfBibitem
\bibitem{Huber:2020vup}
T.~Huber {\em et~al.},
  \ifthenelse{\boolean{articletitles}}{\emph{{Phenomenology of inclusive $
  \Bb\to {X}_s{\mathrm{\ell}}^{+}{\mathrm{\ell}}^{-} $ for the Belle II era}},
  }{}\href{https://doi.org/10.1007/JHEP10(2020)088}{JHEP \textbf{10} (2020)
  088}, \href{http://arxiv.org/abs/2007.04191}{{\normalfont\ttfamily
  arXiv:2007.04191}}\relax
\mciteBstWouldAddEndPuncttrue
\mciteSetBstMidEndSepPunct{\mcitedefaultmidpunct}
{\mcitedefaultendpunct}{\mcitedefaultseppunct}\relax
\EndOfBibitem
\bibitem{Belle:2005fli}
Belle collaboration, M.~Iwasaki {\em et~al.},
  \ifthenelse{\boolean{articletitles}}{\emph{{Improved measurement of the
  electroweak penguin process $B \to X_s l^+ l^-$}},
  }{}\href{https://doi.org/10.1103/PhysRevD.72.092005}{Phys.\ Rev.\
  \textbf{D72} (2005) 092005},
  \href{http://arxiv.org/abs/hep-ex/0503044}{{\normalfont\ttfamily
  arXiv:hep-ex/0503044}}\relax
\mciteBstWouldAddEndPuncttrue
\mciteSetBstMidEndSepPunct{\mcitedefaultmidpunct}
{\mcitedefaultendpunct}{\mcitedefaultseppunct}\relax
\EndOfBibitem
\bibitem{BaBar:2013qry}
BaBar collaboration, J.~P. Lees {\em et~al.},
  \ifthenelse{\boolean{articletitles}}{\emph{{Measurement of the $B \to X_s
  l^+l^-$ branching fraction and search for direct CP violation from a sum of
  exclusive final states}},
  }{}\href{https://doi.org/10.1103/PhysRevLett.112.211802}{Phys.\ Rev.\ Lett.\
  \textbf{112} (2014) 211802},
  \href{http://arxiv.org/abs/1312.5364}{{\normalfont\ttfamily
  arXiv:1312.5364}}\relax
\mciteBstWouldAddEndPuncttrue
\mciteSetBstMidEndSepPunct{\mcitedefaultmidpunct}
{\mcitedefaultendpunct}{\mcitedefaultseppunct}\relax
\EndOfBibitem
\bibitem{Amhis:2021oik}
Y.~Amhis and P.~Owen, \ifthenelse{\boolean{articletitles}}{\emph{{Isospin
  extrapolation as a method to study inclusive $\bar{B} \to X_{s}
  \ell^{+}\ell^{-}$ decays}},
  }{}\href{http://arxiv.org/abs/2106.15943}{{\normalfont\ttfamily
  arXiv:2106.15943}}\relax
\mciteBstWouldAddEndPuncttrue
\mciteSetBstMidEndSepPunct{\mcitedefaultmidpunct}
{\mcitedefaultendpunct}{\mcitedefaultseppunct}\relax
\EndOfBibitem
\bibitem{LHCb:2015hsa}
LHCb collaboration, R.~Aaij {\em et~al.},
  \ifthenelse{\boolean{articletitles}}{\emph{{First measurement of the
  differential branching fraction and $C\!P$ asymmetry of the
  $B^\pm\to\pi^\pm\mu^+\mu^-$ decay}},
  }{}\href{https://doi.org/10.1007/JHEP10(2015)034}{JHEP \textbf{10} (2015)
  034}, \href{http://arxiv.org/abs/1509.00414}{{\normalfont\ttfamily
  arXiv:1509.00414}}\relax
\mciteBstWouldAddEndPuncttrue
\mciteSetBstMidEndSepPunct{\mcitedefaultmidpunct}
{\mcitedefaultendpunct}{\mcitedefaultseppunct}\relax
\EndOfBibitem
\bibitem{Egede:2008uy}
U.~Egede {\em et~al.}, \ifthenelse{\boolean{articletitles}}{\emph{{New
  observables in the decay mode $\bar B_d \to \bar K^{*0} l^+ l^-$}},
  }{}\href{https://doi.org/10.1088/1126-6708/2008/11/032}{JHEP \textbf{11}
  (2008) 032}, \href{http://arxiv.org/abs/0807.2589}{{\normalfont\ttfamily
  arXiv:0807.2589}}\relax
\mciteBstWouldAddEndPuncttrue
\mciteSetBstMidEndSepPunct{\mcitedefaultmidpunct}
{\mcitedefaultendpunct}{\mcitedefaultseppunct}\relax
\EndOfBibitem
\bibitem{LHCb:2013zuf}
LHCb collaboration, R.~Aaij {\em et~al.},
  \ifthenelse{\boolean{articletitles}}{\emph{{Differential branching fraction
  and angular analysis of the decay $B^{0} \to K^{*0} \mu^{+}\mu^{-}$}},
  }{}\href{https://doi.org/10.1007/JHEP08(2013)131}{JHEP \textbf{08} (2013)
  131}, \href{http://arxiv.org/abs/1304.6325}{{\normalfont\ttfamily
  arXiv:1304.6325}}\relax
\mciteBstWouldAddEndPuncttrue
\mciteSetBstMidEndSepPunct{\mcitedefaultmidpunct}
{\mcitedefaultendpunct}{\mcitedefaultseppunct}\relax
\EndOfBibitem
\bibitem{LHCb:2020lmf}
LHCb collaboration, R.~Aaij {\em et~al.},
  \ifthenelse{\boolean{articletitles}}{\emph{{Measurement of \CP-averaged
  observables in the $B^{0}\rightarrow K^{*0}\mu^{+}\mu^{-}$ decay}},
  }{}\href{https://doi.org/10.1103/PhysRevLett.125.011802}{Phys.\ Rev.\ Lett.\
  \textbf{125} (2020) 011802},
  \href{http://arxiv.org/abs/2003.04831}{{\normalfont\ttfamily
  arXiv:2003.04831}}\relax
\mciteBstWouldAddEndPuncttrue
\mciteSetBstMidEndSepPunct{\mcitedefaultmidpunct}
{\mcitedefaultendpunct}{\mcitedefaultseppunct}\relax
\EndOfBibitem
\bibitem{Aaboud:2018krd}
ATLAS collaboration, M.~Aaboud {\em et~al.},
  \ifthenelse{\boolean{articletitles}}{\emph{{Angular analysis of $B^0_d
  \rightarrow K^{*}\mu^+\mu^-$ decays in $pp$ collisions at $\sqrt{s}= 8$ TeV
  with the ATLAS detector}},
  }{}\href{https://doi.org/10.1007/JHEP10(2018)047}{JHEP \textbf{10} (2018)
  047}, \href{http://arxiv.org/abs/1805.04000}{{\normalfont\ttfamily
  arXiv:1805.04000}}\relax
\mciteBstWouldAddEndPuncttrue
\mciteSetBstMidEndSepPunct{\mcitedefaultmidpunct}
{\mcitedefaultendpunct}{\mcitedefaultseppunct}\relax
\EndOfBibitem
\bibitem{Aubert:2006vb}
BaBar collaboration, B.~Aubert {\em et~al.},
  \ifthenelse{\boolean{articletitles}}{\emph{{Measurements of branching
  fractions, rate asymmetries, and angular distributions in the rare decays $\B
  \ra \kaon \ell^+\ell^-$ and \mbox{$\B \ra \Kstar \ell^+\ell^-$}}},
  }{}\href{https://doi.org/10.1103/PhysRevD.73.092001}{Phys.\ Rev.\
  \textbf{D73} (2006) 092001},
  \href{http://arxiv.org/abs/hep-ex/0604007}{{\normalfont\ttfamily
  arXiv:hep-ex/0604007}}\relax
\mciteBstWouldAddEndPuncttrue
\mciteSetBstMidEndSepPunct{\mcitedefaultmidpunct}
{\mcitedefaultendpunct}{\mcitedefaultseppunct}\relax
\EndOfBibitem
\bibitem{Wehle:2016yoi}
Belle collaboration, S.~Wehle {\em et~al.},
  \ifthenelse{\boolean{articletitles}}{\emph{{Lepton-Flavor-Dependent Angular
  Analysis of $B\to K^\ast \ell^+\ell^-$}},
  }{}\href{https://doi.org/10.1103/PhysRevLett.118.111801}{Phys.\ Rev.\ Lett.\
  \textbf{118} (2017) 111801},
  \href{http://arxiv.org/abs/1612.05014}{{\normalfont\ttfamily
  arXiv:1612.05014}}\relax
\mciteBstWouldAddEndPuncttrue
\mciteSetBstMidEndSepPunct{\mcitedefaultmidpunct}
{\mcitedefaultendpunct}{\mcitedefaultseppunct}\relax
\EndOfBibitem
\bibitem{Aaltonen:2011ja}
CDF collaboration, T.~Aaltonen {\em et~al.},
  \ifthenelse{\boolean{articletitles}}{\emph{{Measurements of the angular
  distributions in the decays $\B \to \kaon^{(*)}\mumu$ at CDF}},
  }{}\href{https://doi.org/10.1103/PhysRevLett.108.081807}{Phys.\ Rev.\ Lett.\
  \textbf{108} (2012) 081807},
  \href{http://arxiv.org/abs/1108.0695}{{\normalfont\ttfamily
  arXiv:1108.0695}}\relax
\mciteBstWouldAddEndPuncttrue
\mciteSetBstMidEndSepPunct{\mcitedefaultmidpunct}
{\mcitedefaultendpunct}{\mcitedefaultseppunct}\relax
\EndOfBibitem
\bibitem{Sirunyan:2017dhj}
CMS collaboration, A.~M. Sirunyan {\em et~al.},
  \ifthenelse{\boolean{articletitles}}{\emph{{Measurement of angular parameters
  from the decay $\BdToKstmm$ in proton-proton collisions at $\sqrt{s} = $ 8
  TeV}}, }{}\href{https://doi.org/10.1016/j.physletb.2018.04.030}{Phys.\ Lett.\
   \textbf{B781} (2018) 517},
  \href{http://arxiv.org/abs/1710.02846}{{\normalfont\ttfamily
  arXiv:1710.02846}}\relax
\mciteBstWouldAddEndPuncttrue
\mciteSetBstMidEndSepPunct{\mcitedefaultmidpunct}
{\mcitedefaultendpunct}{\mcitedefaultseppunct}\relax
\EndOfBibitem
\bibitem{Ali:1999mm}
A.~Ali, P.~Ball, L.~T. Handoko, and G.~Hiller,
  \ifthenelse{\boolean{articletitles}}{\emph{{A Comparative study of the decays
  $B \to$ ($K$, $K^{*)} \ell^+ \ell^-$ in standard model and supersymmetric
  theories}}, }{}\href{https://doi.org/10.1103/PhysRevD.61.074024}{Phys.\ Rev.\
   \textbf{D61} (2000) 074024},
  \href{http://arxiv.org/abs/hep-ph/9910221}{{\normalfont\ttfamily
  arXiv:hep-ph/9910221}}\relax
\mciteBstWouldAddEndPuncttrue
\mciteSetBstMidEndSepPunct{\mcitedefaultmidpunct}
{\mcitedefaultendpunct}{\mcitedefaultseppunct}\relax
\EndOfBibitem
\bibitem{Kruger:2005ep}
F.~Kruger and J.~Matias, \ifthenelse{\boolean{articletitles}}{\emph{{Probing
  new physics via the transverse amplitudes of $B^0\to K^{*0} (\to K^- \pi^+)
  l^+l^-$ at large recoil}},
  }{}\href{https://doi.org/10.1103/PhysRevD.71.094009}{Phys.\ Rev.\
  \textbf{D71} (2005) 094009},
  \href{http://arxiv.org/abs/hep-ph/0502060}{{\normalfont\ttfamily
  arXiv:hep-ph/0502060}}\relax
\mciteBstWouldAddEndPuncttrue
\mciteSetBstMidEndSepPunct{\mcitedefaultmidpunct}
{\mcitedefaultendpunct}{\mcitedefaultseppunct}\relax
\EndOfBibitem
\bibitem{Descotes-Genon:2012isb}
S.~Descotes-Genon, J.~Matias, M.~Ramon, and J.~Virto,
  \ifthenelse{\boolean{articletitles}}{\emph{{Implications from clean
  observables for the binned analysis of $B -> K*\mu^+\mu^-$ at large recoil}},
  }{}\href{https://doi.org/10.1007/JHEP01(2013)048}{JHEP \textbf{01} (2013)
  048}, \href{http://arxiv.org/abs/1207.2753}{{\normalfont\ttfamily
  arXiv:1207.2753}}\relax
\mciteBstWouldAddEndPuncttrue
\mciteSetBstMidEndSepPunct{\mcitedefaultmidpunct}
{\mcitedefaultendpunct}{\mcitedefaultseppunct}\relax
\EndOfBibitem
\bibitem{Bharucha:2015bzk}
A.~Bharucha, D.~M. Straub, and R.~Zwicky,
  \ifthenelse{\boolean{articletitles}}{\emph{{$B\to V\ell^+\ell^-$ in the
  Standard Model from light-cone sum rules}},
  }{}\href{https://doi.org/10.1007/JHEP08(2016)098}{JHEP \textbf{08} (2016)
  098}, \href{http://arxiv.org/abs/1503.05534}{{\normalfont\ttfamily
  arXiv:1503.05534}}\relax
\mciteBstWouldAddEndPuncttrue
\mciteSetBstMidEndSepPunct{\mcitedefaultmidpunct}
{\mcitedefaultendpunct}{\mcitedefaultseppunct}\relax
\EndOfBibitem
\bibitem{Horgan:2013hoa}
R.~R. Horgan, Z.~Liu, S.~Meinel, and M.~Wingate,
  \ifthenelse{\boolean{articletitles}}{\emph{{Lattice QCD calculation of form
  factors describing the rare decays $B \to K^* \ell^+ \ell^-$ and $B_s \to
  \phi \ell^+ \ell^-$}},
  }{}\href{https://doi.org/10.1103/PhysRevD.89.094501}{Phys.\ Rev.\
  \textbf{D89} (2014) 094501},
  \href{http://arxiv.org/abs/1310.3722}{{\normalfont\ttfamily
  arXiv:1310.3722}}\relax
\mciteBstWouldAddEndPuncttrue
\mciteSetBstMidEndSepPunct{\mcitedefaultmidpunct}
{\mcitedefaultendpunct}{\mcitedefaultseppunct}\relax
\EndOfBibitem
\bibitem{Kruger:1999xa}
F.~Kruger, L.~M. Sehgal, N.~Sinha, and R.~Sinha,
  \ifthenelse{\boolean{articletitles}}{\emph{{Angular distribution and CP
  asymmetries in the decays $\bar B \to K^- \pi^+ e^- e^+$ and $\bar B \to
  \pi^- \pi^+ e^- e^+$}},
  }{}\href{https://doi.org/10.1103/PhysRevD.61.114028}{Phys.\ Rev.\ D
  \textbf{61} (2000) 114028},
  \href{http://arxiv.org/abs/hep-ph/9907386}{{\normalfont\ttfamily
  arXiv:hep-ph/9907386}}, [Erratum: Phys.Rev.D 63, 019901 (2001)]\relax
\mciteBstWouldAddEndPuncttrue
\mciteSetBstMidEndSepPunct{\mcitedefaultmidpunct}
{\mcitedefaultendpunct}{\mcitedefaultseppunct}\relax
\EndOfBibitem
\bibitem{Descotes-Genon:2013vna}
S.~Descotes-Genon, T.~Hurth, J.~Matias, and J.~Virto,
  \ifthenelse{\boolean{articletitles}}{\emph{{Optimizing the basis of $B\to
  K^*ll$ observables in the full kinematic range}},
  }{}\href{https://doi.org/10.1007/JHEP05(2013)137}{JHEP \textbf{05} (2013)
  137}, \href{http://arxiv.org/abs/1303.5794}{{\normalfont\ttfamily
  arXiv:1303.5794}}\relax
\mciteBstWouldAddEndPuncttrue
\mciteSetBstMidEndSepPunct{\mcitedefaultmidpunct}
{\mcitedefaultendpunct}{\mcitedefaultseppunct}\relax
\EndOfBibitem
\bibitem{Altmannshofer:2008dz}
W.~Altmannshofer {\em et~al.},
  \ifthenelse{\boolean{articletitles}}{\emph{{Symmetries and Asymmetries of $B
  \to K^{*} \mu^{+} \mu^{-}$ Decays in the Standard Model and Beyond}},
  }{}\href{https://doi.org/10.1088/1126-6708/2009/01/019}{JHEP \textbf{01}
  (2009) 019}, \href{http://arxiv.org/abs/0811.1214}{{\normalfont\ttfamily
  arXiv:0811.1214}}\relax
\mciteBstWouldAddEndPuncttrue
\mciteSetBstMidEndSepPunct{\mcitedefaultmidpunct}
{\mcitedefaultendpunct}{\mcitedefaultseppunct}\relax
\EndOfBibitem
\bibitem{Jager:2014rwa}
S.~J{\"a}ger and J.~Martin~Camalich,
  \ifthenelse{\boolean{articletitles}}{\emph{{Reassessing the discovery
  potential of the \mbox{$B \to K^{*} \ell^+\ell^-$} decays in the large-recoil
  region: SM challenges and BSM opportunities}},
  }{}\href{https://doi.org/10.1103/PhysRevD.93.014028}{Phys.\ Rev.\
  \textbf{D93} (2016) 014028},
  \href{http://arxiv.org/abs/1412.3183}{{\normalfont\ttfamily
  arXiv:1412.3183}}\relax
\mciteBstWouldAddEndPuncttrue
\mciteSetBstMidEndSepPunct{\mcitedefaultmidpunct}
{\mcitedefaultendpunct}{\mcitedefaultseppunct}\relax
\EndOfBibitem
\bibitem{Lyon:2014hpa}
J.~Lyon and R.~Zwicky, \ifthenelse{\boolean{articletitles}}{\emph{{Resonances
  gone topsy turvy - the charm of QCD or new physics in \bsll?}},
  }{}\href{http://arxiv.org/abs/1406.0566}{{\normalfont\ttfamily
  arXiv:1406.0566}}\relax
\mciteBstWouldAddEndPuncttrue
\mciteSetBstMidEndSepPunct{\mcitedefaultmidpunct}
{\mcitedefaultendpunct}{\mcitedefaultseppunct}\relax
\EndOfBibitem
\bibitem{Ciuchini:2015qxb}
M.~Ciuchini {\em et~al.}, \ifthenelse{\boolean{articletitles}}{\emph{{$B\to K^*
  \ell^+ \ell^-$ decays at large recoil in the Standard Model: a theoretical
  reappraisal}}, }{}\href{https://doi.org/10.1007/JHEP06(2016)116}{JHEP
  \textbf{06} (2016) 116},
  \href{http://arxiv.org/abs/1512.07157}{{\normalfont\ttfamily
  arXiv:1512.07157}}\relax
\mciteBstWouldAddEndPuncttrue
\mciteSetBstMidEndSepPunct{\mcitedefaultmidpunct}
{\mcitedefaultendpunct}{\mcitedefaultseppunct}\relax
\EndOfBibitem
\bibitem{Bobeth:2017vxj}
C.~Bobeth, M.~Chrzaszcz, D.~van Dyk, and J.~Virto,
  \ifthenelse{\boolean{articletitles}}{\emph{{Long-distance effects in
  $B\rightarrow K^*\ell \ell $ from analyticity}},
  }{}\href{https://doi.org/10.1140/epjc/s10052-018-5918-6}{Eur.\ Phys.\ J.\
  \textbf{C78} (2018) 451},
  \href{http://arxiv.org/abs/1707.07305}{{\normalfont\ttfamily
  arXiv:1707.07305}}\relax
\mciteBstWouldAddEndPuncttrue
\mciteSetBstMidEndSepPunct{\mcitedefaultmidpunct}
{\mcitedefaultendpunct}{\mcitedefaultseppunct}\relax
\EndOfBibitem
\bibitem{Blake:2017fyh}
T.~Blake {\em et~al.}, \ifthenelse{\boolean{articletitles}}{\emph{{An empirical
  model to determine the hadronic resonance contributions to $\overline{B}{} ^0
  \!\rightarrow \overline{K}{} ^{*0} \mu ^+ \mu ^- $ transitions}},
  }{}\href{https://doi.org/10.1140/epjc/s10052-018-5937-3}{Eur.\ Phys.\ J.\
  \textbf{C78} (2018) 453},
  \href{http://arxiv.org/abs/1709.03921}{{\normalfont\ttfamily
  arXiv:1709.03921}}\relax
\mciteBstWouldAddEndPuncttrue
\mciteSetBstMidEndSepPunct{\mcitedefaultmidpunct}
{\mcitedefaultendpunct}{\mcitedefaultseppunct}\relax
\EndOfBibitem
\bibitem{LHCb:2020gog}
LHCb collaboration, R.~Aaij {\em et~al.},
  \ifthenelse{\boolean{articletitles}}{\emph{{Angular Analysis of the
  $B^{+}\rightarrow K^{\ast+}\mu^{+}\mu^{-}$ Decay}},
  }{}\href{https://doi.org/10.1103/PhysRevLett.126.161802}{Phys.\ Rev.\ Lett.\
  \textbf{126} (2021) 161802},
  \href{http://arxiv.org/abs/2012.13241}{{\normalfont\ttfamily
  arXiv:2012.13241}}\relax
\mciteBstWouldAddEndPuncttrue
\mciteSetBstMidEndSepPunct{\mcitedefaultmidpunct}
{\mcitedefaultendpunct}{\mcitedefaultseppunct}\relax
\EndOfBibitem
\bibitem{LHCb:2014auh}
LHCb collaboration, R.~Aaij {\em et~al.},
  \ifthenelse{\boolean{articletitles}}{\emph{{Angular analysis of charged and
  neutral $B \to K \mu^+\mu^-$ decays}},
  }{}\href{https://doi.org/10.1007/JHEP05(2014)082}{JHEP \textbf{05} (2014)
  082}, \href{http://arxiv.org/abs/1403.8045}{{\normalfont\ttfamily
  arXiv:1403.8045}}\relax
\mciteBstWouldAddEndPuncttrue
\mciteSetBstMidEndSepPunct{\mcitedefaultmidpunct}
{\mcitedefaultendpunct}{\mcitedefaultseppunct}\relax
\EndOfBibitem
\bibitem{LHCb:2021xxq}
LHCb collaboration, R.~Aaij {\em et~al.},
  \ifthenelse{\boolean{articletitles}}{\emph{{Angular analysis of the rare
  decay ${B}_s^0 \to \phi\mu^+\mu^-$}},
  }{}\href{https://doi.org/10.1007/JHEP11(2021)043}{JHEP \textbf{11} (2021)
  043}, \href{http://arxiv.org/abs/2107.13428}{{\normalfont\ttfamily
  arXiv:2107.13428}}\relax
\mciteBstWouldAddEndPuncttrue
\mciteSetBstMidEndSepPunct{\mcitedefaultmidpunct}
{\mcitedefaultendpunct}{\mcitedefaultseppunct}\relax
\EndOfBibitem
\bibitem{LHCb:2018jna}
LHCb collaboration, R.~Aaij {\em et~al.},
  \ifthenelse{\boolean{articletitles}}{\emph{{Angular moments of the decay
  $\Lambda_b^0 \rightarrow \Lambda \mu^{+} \mu^{-}$ at low hadronic recoil}},
  }{}\href{https://doi.org/10.1007/JHEP09(2018)146}{JHEP \textbf{09} (2018)
  146}, \href{http://arxiv.org/abs/1808.00264}{{\normalfont\ttfamily
  arXiv:1808.00264}}\relax
\mciteBstWouldAddEndPuncttrue
\mciteSetBstMidEndSepPunct{\mcitedefaultmidpunct}
{\mcitedefaultendpunct}{\mcitedefaultseppunct}\relax
\EndOfBibitem
\bibitem{LHCb:2020dof}
LHCb collaboration, R.~Aaij {\em et~al.},
  \ifthenelse{\boolean{articletitles}}{\emph{{Strong constraints on the $b \to
  s\gamma$ photon polarisation from $B^0 \to K^{*0} e^+ e^-$ decays}},
  }{}\href{https://doi.org/10.1007/JHEP12(2020)081}{JHEP \textbf{12} (2020)
  081}, \href{http://arxiv.org/abs/2010.06011}{{\normalfont\ttfamily
  arXiv:2010.06011}}\relax
\mciteBstWouldAddEndPuncttrue
\mciteSetBstMidEndSepPunct{\mcitedefaultmidpunct}
{\mcitedefaultendpunct}{\mcitedefaultseppunct}\relax
\EndOfBibitem
\bibitem{Capdevila:2016ivx}
B.~Capdevila, S.~Descotes-Genon, J.~Matias, and J.~Virto,
  \ifthenelse{\boolean{articletitles}}{\emph{{Assessing lepton-flavour
  non-universality from $B\to K^*\ell\ell$ angular analyses}},
  }{}\href{https://doi.org/10.1007/JHEP10(2016)075}{JHEP \textbf{10} (2016)
  075}, \href{http://arxiv.org/abs/1605.03156}{{\normalfont\ttfamily
  arXiv:1605.03156}}\relax
\mciteBstWouldAddEndPuncttrue
\mciteSetBstMidEndSepPunct{\mcitedefaultmidpunct}
{\mcitedefaultendpunct}{\mcitedefaultseppunct}\relax
\EndOfBibitem
\bibitem{Belle:2016fev}
Belle collaboration, S.~Wehle {\em et~al.},
  \ifthenelse{\boolean{articletitles}}{\emph{{Lepton-Flavor-Dependent angular
  analysis of $B\to K^\ast \ell^+\ell^-$}},
  }{}\href{https://doi.org/10.1103/PhysRevLett.118.111801}{Phys.\ Rev.\ Lett.\
  \textbf{118} (2017) 111801},
  \href{http://arxiv.org/abs/1612.05014}{{\normalfont\ttfamily
  arXiv:1612.05014}}\relax
\mciteBstWouldAddEndPuncttrue
\mciteSetBstMidEndSepPunct{\mcitedefaultmidpunct}
{\mcitedefaultendpunct}{\mcitedefaultseppunct}\relax
\EndOfBibitem
\bibitem{LHCb:2014mit}
LHCb collaboration, R.~Aaij {\em et~al.},
  \ifthenelse{\boolean{articletitles}}{\emph{{Measurement of $C\!P$ asymmetries
  in the decays $B^0 \rightarrow K^{*0} \mu^+ \mu^-$ and $B^+ \rightarrow K^{+}
  \mu^+ \mu^-$}}, }{}\href{https://doi.org/10.1007/JHEP09(2014)177}{JHEP
  \textbf{09} (2014) 177},
  \href{http://arxiv.org/abs/1408.0978}{{\normalfont\ttfamily
  arXiv:1408.0978}}\relax
\mciteBstWouldAddEndPuncttrue
\mciteSetBstMidEndSepPunct{\mcitedefaultmidpunct}
{\mcitedefaultendpunct}{\mcitedefaultseppunct}\relax
\EndOfBibitem
\bibitem{Benzke:2010tq}
M.~Benzke, S.~J. Lee, M.~Neubert, and G.~Paz,
  \ifthenelse{\boolean{articletitles}}{\emph{{Long-Distance Dominance of the CP
  Asymmetry in $B\to X_{s,d}+\gamma$ Decays}},
  }{}\href{https://doi.org/10.1103/PhysRevLett.106.141801}{Phys.\ Rev.\ Lett.\
  \textbf{106} (2011) 141801},
  \href{http://arxiv.org/abs/1012.3167}{{\normalfont\ttfamily
  arXiv:1012.3167}}\relax
\mciteBstWouldAddEndPuncttrue
\mciteSetBstMidEndSepPunct{\mcitedefaultmidpunct}
{\mcitedefaultendpunct}{\mcitedefaultseppunct}\relax
\EndOfBibitem
\bibitem{BaBar:2014czi}
BaBar collaboration, J.~P. Lees {\em et~al.},
  \ifthenelse{\boolean{articletitles}}{\emph{{Measurements of direct CP
  asymmetries in $B \to X_s\gamma$ decays using sum of exclusive decays}},
  }{}\href{https://doi.org/10.1103/PhysRevD.90.092001}{Phys.\ Rev.\
  \textbf{D90} (2014) 092001},
  \href{http://arxiv.org/abs/1406.0534}{{\normalfont\ttfamily
  arXiv:1406.0534}}\relax
\mciteBstWouldAddEndPuncttrue
\mciteSetBstMidEndSepPunct{\mcitedefaultmidpunct}
{\mcitedefaultendpunct}{\mcitedefaultseppunct}\relax
\EndOfBibitem
\bibitem{Belle:2018iff}
Belle collaboration, S.~Watanuki {\em et~al.},
  \ifthenelse{\boolean{articletitles}}{\emph{{Measurements of isospin asymmetry
  and difference of direct $CP$ asymmetries in inclusive $B \to X_s \gamma$
  decays}}, }{}\href{https://doi.org/10.1103/PhysRevD.99.032012}{Phys.\ Rev.\
  \textbf{D99} (2019) 032012},
  \href{http://arxiv.org/abs/1807.04236}{{\normalfont\ttfamily
  arXiv:1807.04236}}\relax
\mciteBstWouldAddEndPuncttrue
\mciteSetBstMidEndSepPunct{\mcitedefaultmidpunct}
{\mcitedefaultendpunct}{\mcitedefaultseppunct}\relax
\EndOfBibitem
\bibitem{BaBar:2008okc}
BaBar collaboration, B.~Aubert {\em et~al.},
  \ifthenelse{\boolean{articletitles}}{\emph{{Measurement of Time-Dependent CP
  Asymmetry in $B^0 \to K^0_{S} \pi^0 \gamma$ Decays}},
  }{}\href{https://doi.org/10.1103/PhysRevD.78.071102}{Phys.\ Rev.\ D
  \textbf{78} (2008) 071102},
  \href{http://arxiv.org/abs/0807.3103}{{\normalfont\ttfamily
  arXiv:0807.3103}}\relax
\mciteBstWouldAddEndPuncttrue
\mciteSetBstMidEndSepPunct{\mcitedefaultmidpunct}
{\mcitedefaultendpunct}{\mcitedefaultseppunct}\relax
\EndOfBibitem
\bibitem{Belle:2007kjy}
Belle collaboration, Y.~Ushiroda {\em et~al.},
  \ifthenelse{\boolean{articletitles}}{\emph{{Time-dependent CP-violating
  asymmetry in B0 ---\ensuremath{>} rho0 gamma decays}},
  }{}\href{https://doi.org/10.1103/PhysRevLett.100.021602}{Phys.\ Rev.\ Lett.\
  \textbf{100} (2008) 021602},
  \href{http://arxiv.org/abs/0709.2769}{{\normalfont\ttfamily
  arXiv:0709.2769}}\relax
\mciteBstWouldAddEndPuncttrue
\mciteSetBstMidEndSepPunct{\mcitedefaultmidpunct}
{\mcitedefaultendpunct}{\mcitedefaultseppunct}\relax
\EndOfBibitem
\bibitem{LHCb:2019vks}
LHCb collaboration, R.~Aaij {\em et~al.},
  \ifthenelse{\boolean{articletitles}}{\emph{{Measurement of $CP$-violating and
  mixing-induced observables in $B^0 \to \phi\gamma$ decays}},
  }{}\href{https://doi.org/10.1103/PhysRevLett.123.081802}{Phys.\ Rev.\ Lett.\
  \textbf{123} (2019) 081802},
  \href{http://arxiv.org/abs/1905.06284}{{\normalfont\ttfamily
  arXiv:1905.06284}}\relax
\mciteBstWouldAddEndPuncttrue
\mciteSetBstMidEndSepPunct{\mcitedefaultmidpunct}
{\mcitedefaultendpunct}{\mcitedefaultseppunct}\relax
\EndOfBibitem
\bibitem{Hiller:2003js}
G.~Hiller and F.~Kr\"uger, \ifthenelse{\boolean{articletitles}}{\emph{{More
  model-independent analysis of \mbox{$b\rightarrow s$} processes}},
  }{}\href{https://doi.org/10.1103/PhysRevD.69.074020}{Phys.\ Rev.\
  \textbf{D69} (2004) 074020},
  \href{http://arxiv.org/abs/hep-ph/0310219}{{\normalfont\ttfamily
  arXiv:hep-ph/0310219}}\relax
\mciteBstWouldAddEndPuncttrue
\mciteSetBstMidEndSepPunct{\mcitedefaultmidpunct}
{\mcitedefaultendpunct}{\mcitedefaultseppunct}\relax
\EndOfBibitem
\bibitem{LHCb-PAPER-2021-004}
LHCb collaboration, R.~Aaij {\em et~al.},
  \ifthenelse{\boolean{articletitles}}{\emph{{Test of lepton universality in
  beauty-quark decays}},
  }{}\href{http://arxiv.org/abs/2103.11769}{{\normalfont\ttfamily
  arXiv:2103.11769}}\relax
\mciteBstWouldAddEndPuncttrue
\mciteSetBstMidEndSepPunct{\mcitedefaultmidpunct}
{\mcitedefaultendpunct}{\mcitedefaultseppunct}\relax
\EndOfBibitem
\bibitem{Alguero:2021anc}
M.~Alguer\'o {\em et~al.}, \ifthenelse{\boolean{articletitles}}{\emph{{$b\to
  s\ell\ell$ global fits after Moriond 2021 results}},
  }{}\href{http://arxiv.org/abs/2104.08921}{{\normalfont\ttfamily
  arXiv:2104.08921}}\relax
\mciteBstWouldAddEndPuncttrue
\mciteSetBstMidEndSepPunct{\mcitedefaultmidpunct}
{\mcitedefaultendpunct}{\mcitedefaultseppunct}\relax
\EndOfBibitem
\bibitem{Altmannshofer:2021qrr}
W.~Altmannshofer and P.~Stangl, \ifthenelse{\boolean{articletitles}}{\emph{{New
  physics in rare B decays after Moriond 2021}},
  }{}\href{https://doi.org/10.1140/epjc/s10052-021-09725-1}{Eur.\ Phys.\ J.\
  \textbf{C81} (2021) 952},
  \href{http://arxiv.org/abs/2103.13370}{{\normalfont\ttfamily
  arXiv:2103.13370}}\relax
\mciteBstWouldAddEndPuncttrue
\mciteSetBstMidEndSepPunct{\mcitedefaultmidpunct}
{\mcitedefaultendpunct}{\mcitedefaultseppunct}\relax
\EndOfBibitem
\bibitem{Ciuchini:2020gvn}
M.~Ciuchini {\em et~al.}, \ifthenelse{\boolean{articletitles}}{\emph{{Lessons
  from the $B^{0,+}\to K^{*0,+}\mu^+\mu^-$ angular analyses}},
  }{}\href{https://doi.org/10.1103/PhysRevD.103.015030}{Phys.\ Rev.\
  \textbf{D103} (2021) 015030},
  \href{http://arxiv.org/abs/2011.01212}{{\normalfont\ttfamily
  arXiv:2011.01212}}\relax
\mciteBstWouldAddEndPuncttrue
\mciteSetBstMidEndSepPunct{\mcitedefaultmidpunct}
{\mcitedefaultendpunct}{\mcitedefaultseppunct}\relax
\EndOfBibitem
\bibitem{Hurth:2021nsi}
T.~Hurth, F.~Mahmoudi, D.~M. Santos, and S.~Neshatpour,
  \ifthenelse{\boolean{articletitles}}{\emph{{More indications for lepton
  nonuniversality in $b \to s \ell^+ \ell^-$}},
  }{}\href{http://arxiv.org/abs/2104.10058}{{\normalfont\ttfamily
  arXiv:2104.10058}}\relax
\mciteBstWouldAddEndPuncttrue
\mciteSetBstMidEndSepPunct{\mcitedefaultmidpunct}
{\mcitedefaultendpunct}{\mcitedefaultseppunct}\relax
\EndOfBibitem
\bibitem{Belle-II:2018jsg}
Belle-II collaboration, W.~Altmannshofer {\em et~al.},
  \ifthenelse{\boolean{articletitles}}{\emph{{The Belle II Physics Book}},
  }{}\href{https://doi.org/10.1093/ptep/ptz106}{PTEP \textbf{2019} (2019)
  123C01}, \href{http://arxiv.org/abs/1808.10567}{{\normalfont\ttfamily
  arXiv:1808.10567}}, [Erratum: PTEP 2020, 029201 (2020)]\relax
\mciteBstWouldAddEndPuncttrue
\mciteSetBstMidEndSepPunct{\mcitedefaultmidpunct}
{\mcitedefaultendpunct}{\mcitedefaultseppunct}\relax
\EndOfBibitem
\bibitem{Kamenik:2017ghi}
J.~F. Kamenik, S.~Monteil, A.~Semkiv, and L.~V. Silva,
  \ifthenelse{\boolean{articletitles}}{\emph{{Lepton polarization asymmetries
  in rare semi-tauonic $ b \rightarrow s $ exclusive decays at FCC-$ee$}},
  }{}\href{https://doi.org/10.1140/epjc/s10052-017-5272-0}{Eur.\ Phys.\ J.\
  \textbf{C77} (2017) 701},
  \href{http://arxiv.org/abs/1705.11106}{{\normalfont\ttfamily
  arXiv:1705.11106}}\relax
\mciteBstWouldAddEndPuncttrue
\mciteSetBstMidEndSepPunct{\mcitedefaultmidpunct}
{\mcitedefaultendpunct}{\mcitedefaultseppunct}\relax
\EndOfBibitem
\bibitem{ATL-PHYS-PUB-2018-005}
ATLAS Collaboration,  \ifthenelse{\boolean{articletitles}}{\emph{{Prospects for
  the ${\cal B}(B^0_{(s)} \to \mu^+ \mu^-)$ measurements with the ATLAS
  detector in the Run 2 and HL-LHC data campaigns}}}{},
  \href{https://cds.cern.ch/record/2317211}{ATL-PHYS-PUB-2018-005}, CERN,
  Geneva, 2018\relax
\mciteBstWouldAddEndPuncttrue
\mciteSetBstMidEndSepPunct{\mcitedefaultmidpunct}
{\mcitedefaultendpunct}{\mcitedefaultseppunct}\relax
\EndOfBibitem
\bibitem{CMS-PAS-FTR-18-013}
CMS Collaboration,  \ifthenelse{\boolean{articletitles}}{\emph{{Measurement of
  rare $B \to \mu^+\mu^-$ decays with the Phase-2 upgraded CMS detector at the
  HL-LHC}}}{}, \href{https://cds.cern.ch/record/2650545}{CMS-PAS-FTR-18-013},
  CERN, Geneva, 2018\relax
\mciteBstWouldAddEndPuncttrue
\mciteSetBstMidEndSepPunct{\mcitedefaultmidpunct}
{\mcitedefaultendpunct}{\mcitedefaultseppunct}\relax
\EndOfBibitem
\end{mcitethebibliography}

\end{document}